\documentclass[useAMS,usenatbib]{mn2e}
\usepackage{graphicx}
\usepackage{dcolumn}
\usepackage{bm}
\usepackage{epsfig}
\usepackage{aas_macros}
\usepackage{fixltx2e}
\usepackage{amsmath}
\usepackage{url}
\newcommand{\LCDM}{$\Lambda$CDM}

\newcommand{\mum}{\mbox{$\mu$m}}

\def\lesssim{\lower.5ex\hbox{$\; \buildrel < \over \sim \;$}}
\def\gtrsim{\lower.5ex\hbox{$\; \buildrel > \over \sim \;$}}
\title[Modeling of the EBL and gamma-ray spectra]{Semi-analytic modeling of the EBL and consequences for extragalactic gamma-ray spectra}

\author[R.~C.~Gilmore, R.~S. Somerville, J.~R.~Primack, A.~Dom\'{i}nguez]{Rudy C. Gilmore$^{1,2}$\thanks{E-mail: rgilmore@physics.ucsc.edu}, Rachel S. Somerville$^{3,4}$, Joel R. Primack$^{2}$, \newauthor Alberto Dom\'{i}nguez$^{5,6,7,8}$\\
$^{1}$ SISSA, via Bonomea 265, 34136 Trieste, Italy \\
$^{2}$ University of California, Santa Cruz, CA 95064 USA\\
$^{3}$ Space Telescope Science Institute, Baltimore, MD 21218\\
$^{4}$ Department of Physics and Astronomy, Johns Hopkins University, Baltimore, MD 21218\\
$^{5}$ Visiting researcher at the Santa Cruz Institute for Particle Physics (SCIPP), University of California, Santa Cruz, CA 95064, USA\\
$^{6}$ Instituto de Astrof\'{i}sica de Andaluc\'{i}a, CSIC, Apdo. Correos 3004, E-18080 Granada, Spain\\
$^{7}$ Departamento de F\'{i}sica At\'{o}mica, Molecular y Nuclear, Universidad de Sevilla, Apdo. Correos 1065, E-41080 Sevilla, Spain\\
$^{8}$ Department of Physics and Astronomy, University of California, Riverside, CA 92521, USA
}

\begin{document}
\date{\today}

\pagerange{\pageref{firstpage}--\pageref{lastpage}} \pubyear{2011}

\maketitle
\label{firstpage}

\begin{abstract}

Attenuation of high-energy gamma rays by pair-production with UV,
optical and IR extragalactic background light (EBL) photons provides a
link between the history of galaxy formation and high-energy
astrophysics.  We present results from our latest semi-analytic models
(SAMs), which employ the main ingredients thought to be important to
galaxy formation and evolution, as well as an improved model for
reprocessing of starlight by dust to mid- and far-IR wavelengths.
These SAMs are based upon a $\Lambda$CDM hierarchical structural
formation scenario, and are successful in reproducing a large variety
of observational constraints such as number counts, luminosity and
mass functions, and color bimodality.  Our fiducial model is based
upon a WMAP5 cosmology, and treats dust emission using empirical templates.  This model
predicts a background flux considerably lower than optical and near
-IR measurements that rely on subtraction of zodiacal and galactic
foregrounds, and near the lower bounds set by number counts of
resolvable sources at a large number of wavelengths.  We also show the
results of varying cosmological parameters and dust attenuation model used in our SAM.  For each EBL prediction, we show how the optical depth
due to electron-positron pair--production is affected by redshift and gamma-ray energy,
and the effect of gamma-ray absorption on the spectra of a variety of
extragalactic sources.  We conclude with a discussion of the
implications of our work, comparisons to other models and key
measurements of the EBL and a discussion of how the burgeoning science
of gamma-ray astronomy will continue to help constrain cosmology.  The
low EBL flux predicted by our fiducial model suggests an optimistic
future for further studies of distant gamma-ray sources.
\end{abstract}

\section{Introduction}
The extragalactic background light (EBL) is the totality of light
emitted by stars and active galactic nuclei (AGN) over the lifetime of
the universe.  Today, this pervasive photon background consists of
light emitted at all epochs, modified by redshifting and dilution due
to the expansion of the universe.  The bulk of the EBL occurs at
wavelengths from the near-UV to the far-IR.  In the UV, optical, and
near-IR most of the EBL is due to direct starlight, as well as a
subdominant contribution from active galactic nuclei (AGN) \citep{schirber&bullock03}.  From the mid-IR to submillimeter wavelengths, the EBL consists of reemitted light from dust particles, including both continuum thermal
radiation and line emission from polycyclic aromatic hydrocarbons
(PAH) molecules \citep{lagache05}.  The background at longer
wavelengths is dominated by the cosmic microwave background, while
shortward of the Lyman limit the background flux decreases rapidly due
to attenuation by neutral hydrogen in stellar atmospheres and the
interstellar and intergalactic media.

Because the production of the EBL is directly linked to the star
formation history of the universe, limits on the EBL can be used to
provide constraints on the history of galaxy formation and evolution.
Observations of the extragalactic sky brightness can constrain the
local background, however they do not provide information about
evolution of the background with redshift.  Direct sky photometry has
been attempted with a number of instruments, most notably DIRBE and
FIRAS, and also ground-based optical telescopes (e.g. \citep{mattila11}), but these type of measurements are subject to considerable
uncertainties due to the large foreground sources that must be
subtracted \citep{hauser&dwek01}.  Integration over discrete sources
seen in galaxy surveys (e.g. \citealp{madau00,keenan10}) is another way to estimate the EBL, but one
that in principle can only provide a lower limit due to the
possibility of unseen sources beyond the magnitude limits of the
survey instrument or underestimation of true total luminosity of
galaxies due to light in the faint outskirts (e.g. \citealp{bernstein07}).  
Observations with highly sensitive satellite instruments have provided us with EBL lower limits from galaxy number counts across wide wavelength ranges.

High-energy gamma rays can interact with EBL photons in electron-positron pair-production interactions \citep{nikishov62,gould&schreder67,jelley66}.  By effectively removing these gamma rays from view, this process has the potential to alter the observed spectra of extragalactic high-energy sources, and increasingly occlude those at higher redshifts.  The rapid development of ground-based gamma-ray astronomy in the past 20 years has led to a number of attempts -- e.g. \citet{dwek&krennrich05,aharonian06,mazin&raue07,albert08} -- to constrain the EBL based on modification to gamma-ray spectra, a method that can provide a measurement of the EBL that is independent of direct observation.  In principle, the cosmological history of the EBL could be reconstructed by comparing observations of high-energy sources at different redshifts to their intrinsic spectra.  Unfortunately, the emission mechanisms and intrinsic spectra of GeV and TeV sources are still poorly understood.

Understanding how the EBL is produced and how its spectral energy distribution (SED) evolves in redshift requires an understanding of the sources responsible for its production.  This has been attempted by different authors using a variety of techniques.  As enumerated in \citet{dominguez11} (D11), calculations of the EBL fall into four general categories: {\bf i)} forward evolution beginning with initial cosmological conditions, such as the semi-analytic models used in this work; {\bf ii)} backwards evolution of the well-constrained present-day galaxy emissivity according to some prescription; {\bf iii)}  evolution of galaxy properties that are inferred over some range in wavelength; {\bf iv)} direct observation of evolution in galaxy properties over the redshifts providing the major contribution to the background light, a category which describes the empirical method developed in D11.

The last two of these have become much more powerful techniques in recent years due to large-scale surveys by ground- and space-based instruments, especially at UV and IR wavelengths, where a great deal of progress has taken place in the last decade.  Some of the first models to account for EBL production by the evolving galaxy population were \citet{madau98} and \citet{franceschini00}, using HST and ISO data, respectively, and \citet{pei99}, who looked at chemical enrichment data in Ly$\alpha$ systems.  A two-part paper series by T. Kneiske and collaborators \citep{kneiske02,kneiske04} computed the EBL and subsequently predictions for attenuation of gamma-ray sources based on a parametrization of the SFR density.  These models separately include the contribution of the LIRG/ULIRG population.  An update to this work by \citet{kneiske&dole10} attempts to create a model using a similar method that produces a minimal background, and a similar method was employed in \citet{razzaque09} and \citet{finke10}.   The aforementioned work of D11 used observed evolution up to $z=4$ in the K-band luminosity function combined with the evolving distribution of 25 galaxy SED types from a multiwavelength survey of galaxies to estimate the EBL and its evolution.

Other authors have used backward evolution models to predict the EBL.
These calculations begin with the present day galaxy luminosity
function and attempt to trace this function backwards in time by
assuming a functional form for the redshift evolution.  In
\citet{malkan&stecker98,malkan&stecker01}, IR luminosity functions
from IRAS were extrapolated backwards in redshift using power law
functions.  The model of \citet{stecker06} updated this work and
computed the EBL below the Lyman limit (13.6 eV) for two different
cases of stellar evolution.  The model of \citet{rowanrobinson01} also
utilized a 60 $\mu$m evolving luminosity function, and a
four-component spectral model for IR and optical emission.  One
potential problem with this method is that it has difficulty
accounting for the emissivity contribution of merger-triggered
starbursts, believed to contribute an increasing fraction of the SFR
density and IR emissivity with increasing redshift.   \citet{franceschini01} made an attempt to account for this starburst phase in a backwards evolution model.  \citet{franceschini08} published a sophisticated model using observed luminosity functions, and used it to calculate the EBL and gamma-ray
attenuation.  This model uses evolving luminosity functions in the
near-IR up to $z=1.4$ for two different galaxy populations (spiral and
spheroidal) and local luminosity functions for the
irregular/starbursting population, combined with synthetic SEDs to
find the total emissivity.

In forward evolution scenarios such as semi-analytic models (SAMs), predictions for the evolution of galaxy emissivities are made by beginning from the universe in its primordial state and simulating the process of galaxy formation.  This is considerably more involved and challenging than the other methods of estimating the EBL, but can provide a degree of insight into the fundamental astrophysics processes that determine the emissivity that is lacking in other approaches.  Semi-analytic models of structure formation based on cold-dark matter (CDM) merger trees have been used in several papers by our group to predict the EBL.  \citet{primack99} predicted the EBL using the SAM described in \citet{somerville&primack99} and \citet{somerville01}.  Later work included improved treatment of absorption and reemission of starlight by dust and updated cosmological data \citep{primack05,primack08}.  

In a companion paper to this work, \citet{sgpd11}, which we will
hereafter refer to as SGPD11, we present a new semi-analytic model
based on galaxy formation in a WMAP5 cosmology.  This model, which
will be summarized in the next section, incorporates the physical
processes thought to be most important in determining the evolution of
these systems.  From the luminosity density calculated in this model,
we have predicted the evolving EBL out to high redshift.  In this
paper, we address the topic of gamma-ray attenuation and show how our
estimated EBL affects high-energy observations of extragalactic
sources, and discuss the ability of gamma-ray telescopes to explore
the distant universe.  We also present a comprehensive comparison to the predictions for EBL and gamma-ray opacity that have been proposed by a number of recent authors.  Our current work is an update to
\citet{primack08}, which presents an earlier stage of our results
using a `concordance cosmology' (C\LCDM), with parameters largely
consistent with WMAP1.  This work is also closely related and
complementary to \citet{gilmoreUV}, which used the C\LCDM~model as the
basis for a prediction of the UV background radiation out to high
redshift, and therefore emphasized the calculation of optical depths
for gamma-rays below 200 GeV. In that paper we included contributions
to the UV emissivity from quasars, as well as an account of the
attenuation of ionizing radiation escaping from galaxies and
processing by neutral hydrogen in the IGM using a radiative transfer
code.

In the following Section, we briefly review the key elements of the semi-analytic model presented in SGPD11.  Results are presented in Section \ref{sec:results}, beginning with a review of key results related to the evolving background radiation from SGPD11 in \ref{subsec:astrores}.  In Section \ref{subsec:gammaatt}, we show how the population of photons in our calculated EBL impacts observations of extragalactic gamma-ray sources through pair-production interactions.  Section \ref{subsec:blazars} deals with the comparison between our predicted gamma-ray optical depths and observations of VHE blazars, and constraints on the EBL that other authors have derived using high-energy data.  In Section \ref{sec:comparison}, we compare our EBL model with several others that have been proposed using a variety of techniques in recent years.  We conclude in Section \ref{sec:ebldisc} with a summary of results and a discussion of how current and future high-energy observations will continue to constrain the EBL.

\section{Model}
\label{sec:model}

This section summarizes the semi-analytic model that is used to
predict the EBL in this work.  This model is based upon the models
that were first presented in \citet{somerville&primack99} and
\citet{somerville01}, with significant new ingredients as described in
\citet{somerville08} (S08). Readers should refer to SGPD11, as well as
S08, for details.

\subsection{Galaxy formation}
We assume a standard $\Lambda$CDM universe and a Chabrier \citep{chabrier03} stellar initial mass function (IMF) that does not evolve in redshift.  The model presented in this work
uses cosmological parameters based on WMAP5, including a power
spectrum normalization of $\sigma_8 = 0.82$, a value that is
intermediate between the previous findings of WMAP1 and WMAP3.  This
value is within $1\sigma$ of the recently-published value from WMAP7
of $0.809\pm0.024$ \citep{komatsu11}.  The SAMs used here are based on
Monte Carlo realizations of dark matter halo merger histories
calculated using the modified Press-Schechter \citep{sheth&tormen99}
and extended Press-Schechter methods.

Star formation occurs when gas is accreted by the galaxy and becomes
available after cooling via atomic processes.  
Feedback from supernovae can heat and eject the cold gas within the
galaxy.  This gas will either be deposited in the hot reservoir
connected with the dark matter halo, or returned to the IGM, depending
on the wind velocity relative to the virial velocity of the halo.  
Star formation in our model occurs in two regimes, quiescent star formation in isolated galaxies and merger-driven starbursts.  The former is treated using a recipe based on the empirical Schmidt-Kennicutt relation \citep{kennicutt89,kennicutt98}.  
Mergers drive gas deep into galactic nuclei, fueling black hole growth which power AGN-driven winds. Supermassive black holes can also produce radio jets that heat the hot halo gas and may eventually shut off cooling and eventually lead to a cessation of star formation. This ``quenching'' of star formation tends to occur in massive galaxies, which are able to build massive BH, and which are accreting gas from hot, tenuous halos rather than via cold dense filaments.

The chemical enrichment and star formation history of each galaxy are
used to predict the total emission spectrum.  We have adopted the
\citet{bruzual&charlot03} stellar population models in this work.

\subsection{Dust extinction and reemission}

Light emitted by stars can be absorbed and reemitted by dust.  In the
SAM, dust is modeled as a two-component distribution, using a modified
version of the prescription of \citet{charlot&fall00}, which treats
separately the dense dust in giant molecular clouds that contain
newborn stars and the much more diffuse cirrus in the interstellar
medium (ISM).   Extinction from the ISM component is proportional to the density and metallicity of cold dust, and a slab geometry is assumed.
Stars younger than $10^7$ yr are enshrouded in a cloud of dust with
optical depth $\tau_{\mathrm{BC,V}}=\mu_{\mathrm{BC}}\, \tau_{V,0}$,
where $\mu_{\mathrm{BC}}=3$ and $\tau_{V,0}$ is the face-on, V-band extinction of the ISM component.  To calculate extinction at other
wavelengths, we have assumed a starburst attenuation curve
\citep{calzetti:00} for the diffuse dust component and a power-law
extinction curve $A_{\lambda}\propto(\lambda/5500 \, \mbox{\AA})^n$,
with $n=0.7$, for the birth clouds \citep{charlot&fall00}.  

We consider two possible normalizations for the extinction recipes.  In
our `WMAP5+fixed' model, parameters are constant at all redshifts, and are
adjusted to match observed relations between UV and IR luminosity for
nearby galaxies.  However, as discussed in SGPD11, we have found that this model has
difficulty reproducing luminosity functions at higher redshift in the
UV- and optical bands.  Motivated by this finding, we have created an
`evolving' model with redshift-dependent parameters tuned to match the
observed UV- and optical luminosity functions at all redshifts where
they have been measured.  In this model, total dust extinction is scaled by a factor $(1+z)^{-1}$ at non-zero redshifts, and the opacity and lifetime of molecular clouds is scaled by a factor $z^{-1}$ above redshift 1.  Because this evolving dust model is found to be more successful at matching high-redshift data, we favor this model and will refer to it as the `Fiducial' variant in the next section.

The reemission of IR light by the dust due to thermal and PAH emission
is estimated in our model using templates that describe the spectra of
galaxies from the mid-IR to submillimeter as a function of the total
IR luminosity, and are based on observations of galaxies in the local
universe.  Energy absorbed by dust from direct starlight is
redistributed in the infrared according to a prescribed SED.  These
templates are embedded in our semi-analytic model, and account for
emission at wavelengths from a few microns to the sub-millimeter,
including the emission and absorption lines appearing in the PAH
region.

The dust emission templates we have used here are described in
\citet[][R09]{rieke09}, and are based on observations of 11 local
LIRGS and ULIRGS combined with lower luminosity local systems.
A comparison of these templates with those of \citet{devriendt99,devriendt00}, which were used in \citet{primack01}, \citet{primack05}, and \citet{primack08}, is available in SGPD11.   Being observationally-based, these templates suffer to some extent from starlight contamination at short wavelengths, and there is also a discontinuity in the galactic SEDs where the templates are joined with our stellar synthesis models, at about 4 microns.  In the wavelength range of 2 to 5 microns we have attempted to compensate for these problems by fitting a power-law extrapolation to the templates. The overall effect of this change to the integrated background light that we will present in the next section is minimal.

\section{Results}
\label{sec:results}

\subsection{Overview of astrophysical results}
\label{subsec:astrores}

In this section, we present results for the evolving background light predicted by our semi-analytic models.  In addition to our fiducial model, which uses a WMAP5 cosmology combined with the evolving dust model and R09 templates described in the last section, we will also show results using employing the `fixed' dust absorption model to facilitate comparison with our older work.  Results from our C\LCDM~model will also be shown when relevant.   As in the previous section, much of the material here is a review of key results from SGPD11, and readers should refer to this work for further detail.  We will also compare with the work of \citet{dominguez11} (D11), a more observationally-driven model that provides a useful contrast to our theoretical approach.

Predictions for the evolving EBL and gamma-ray opacity in the fiducial and fixed models are available online in tabulated form\footnote{\url{http://physics.ucsc.edu/~joel/EBLdata-Gilmore2012/}}, and can also be found bundled with the ArXiv source of this paper.

\subsubsection{Star formation and luminosity density history}

\begin{figure}
\centering
\psfig{file=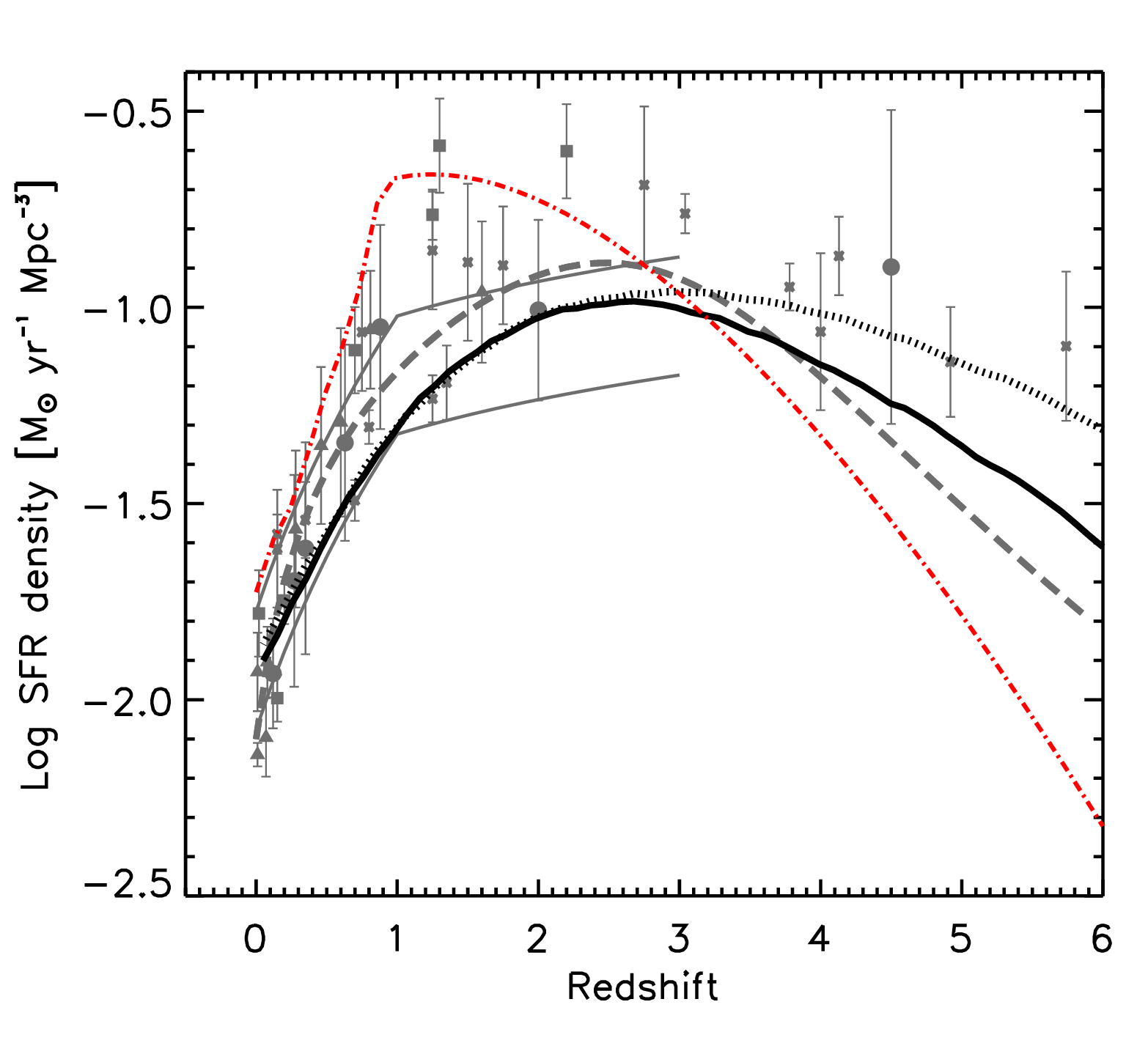,width=7.5cm}
\caption{The star-formation rate density in each of the two SAMs over cosmic time.  The solid and dotted black curves are the predictions from our WMAP5 and C\LCDM~models, respectively.  The dash-dotted red line is the SFR density inferred from the model of D11, which has been converted to a Chabrier IMF.  Grey data points are from a compilation presented in S08; and the dashed line is the estimate of \citet{hopkins&beacom06}.}
\label{fig:sfr}
\end{figure}

The global star formation rate density arising in our WMAP5 and
C\LCDM~models is shown in Figure \ref{fig:sfr}, along with
observational estimates obtained using a variety of tracers.  
Our model makes predictions that are in
agreement with the bulk of data for $z<1$, and tend to be slightly
lower than observed at $z\sim 2$.  All measured star-formation rates
are subject to significant uncertainties, as seen in the scatter in
results for the plotted data.  Uncertainties in dust extinction impact
all results relying on UV luminosity.  Measurements of H$\alpha$ and
other spectral lines must take into account extinction as well as
metallicity effects.  Other authors have attempted to measure
star-formation rates based on 24 $\mu$m and other mid-IR observations
of warm dust.  These results can be affected by AGN contamination, as
well as PAH features that move in and out of the instrument bandpass
with changing redshift.  All of these uncertainties grow with
increasing redshift, where our knowledge of dust distribution and
galaxy SEDs becomes less reliable.  The star-formation rate density inferred from the UV and IR luminosity densities of D11 is also shown here.  The larger value predicted in this work relative to ours is due to the considerably higher far-IR emission in D11; predictions in the optical and near-IR are similar.  Above redshift 1, star-formation rate density predictions from D11 are affected by the assumption made about the evolution of different galaxy spectral types.

\begin{figure*}
\psfig{file=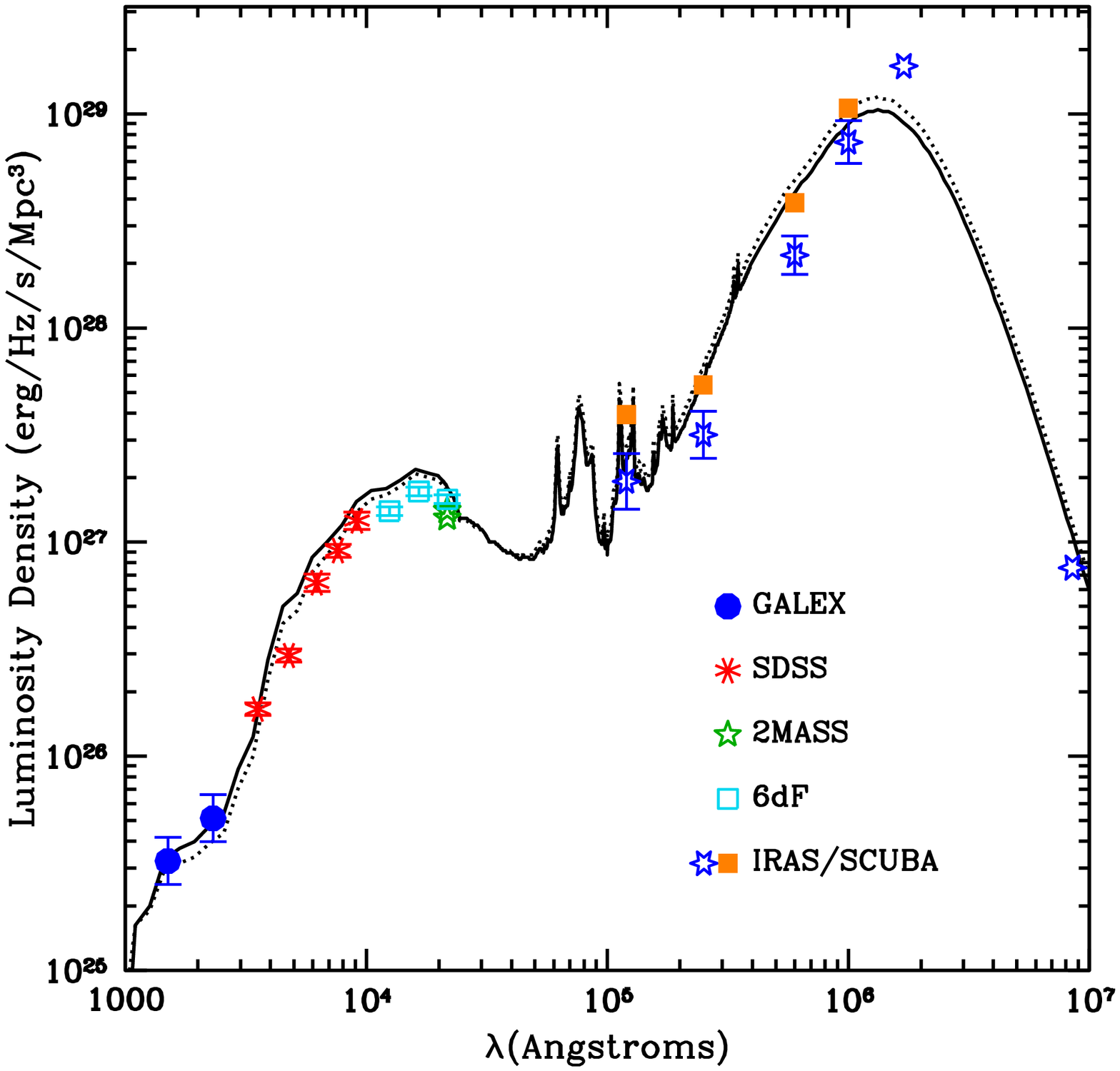,width=8cm}
\psfig{file=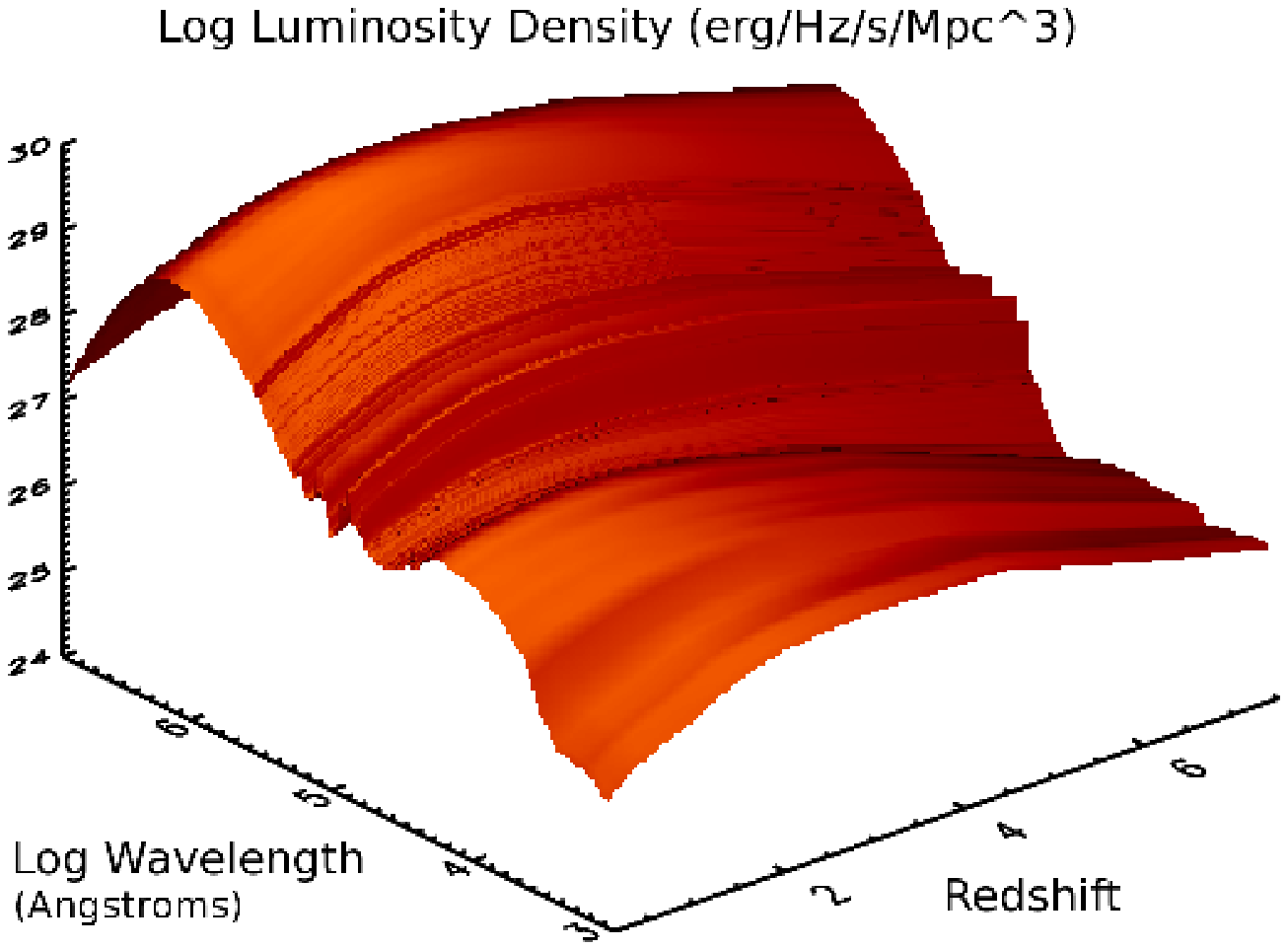,width=8cm}
\caption{{\bf Left:} The luminosity density of the local universe.  The solid black line is the WMAP5 model, and the dotted line is the C\LCDM~model.  Data at a number of wavelengths is shown from GALEX (blue circles), SDSS (red stars; \citealp{md&prada09}), 6dF (light blue squares; \citealp{jones06}), 2MASS  (green stars; \citealp{cole01}, and \citealp{bell03}).  In the mid- and far-IR, the orange squares are from IRAS \citep{soifer91}, while blue stars are from an analysis of local emissivity using data from IRAS, ISO, and SCUBA \citep{takeuchi01}.  {\bf Right:} Three-dimensional representation of the evolution of the luminosity density in our WMAP5 model as a function of wavelength and redshift.}
\label{fig:ld3D}
\end{figure*}

The luminosity density in our models is predicted by summing over the
emission from individual galaxies. Results for total galactic emissivity as a function of wavelength and
redshift are shown in the right-hand panel of Figure \ref{fig:ld3D}.  In
the left panel we show predicted luminosity density in the local
universe, compared to constraints at a number of wavelengths.  The
local luminosity density has been extremely well-measured in the
optical and near-IR by large-scale surveys such as SDSS and 2MASS, and
this provides a strong constraint on any model of the galaxy
population.  At longer wavelengths, we show how our models fit the
local data at IR wavelengths, including data from IRAS and SCUBA.

\begin{figure*}
\psfig{file=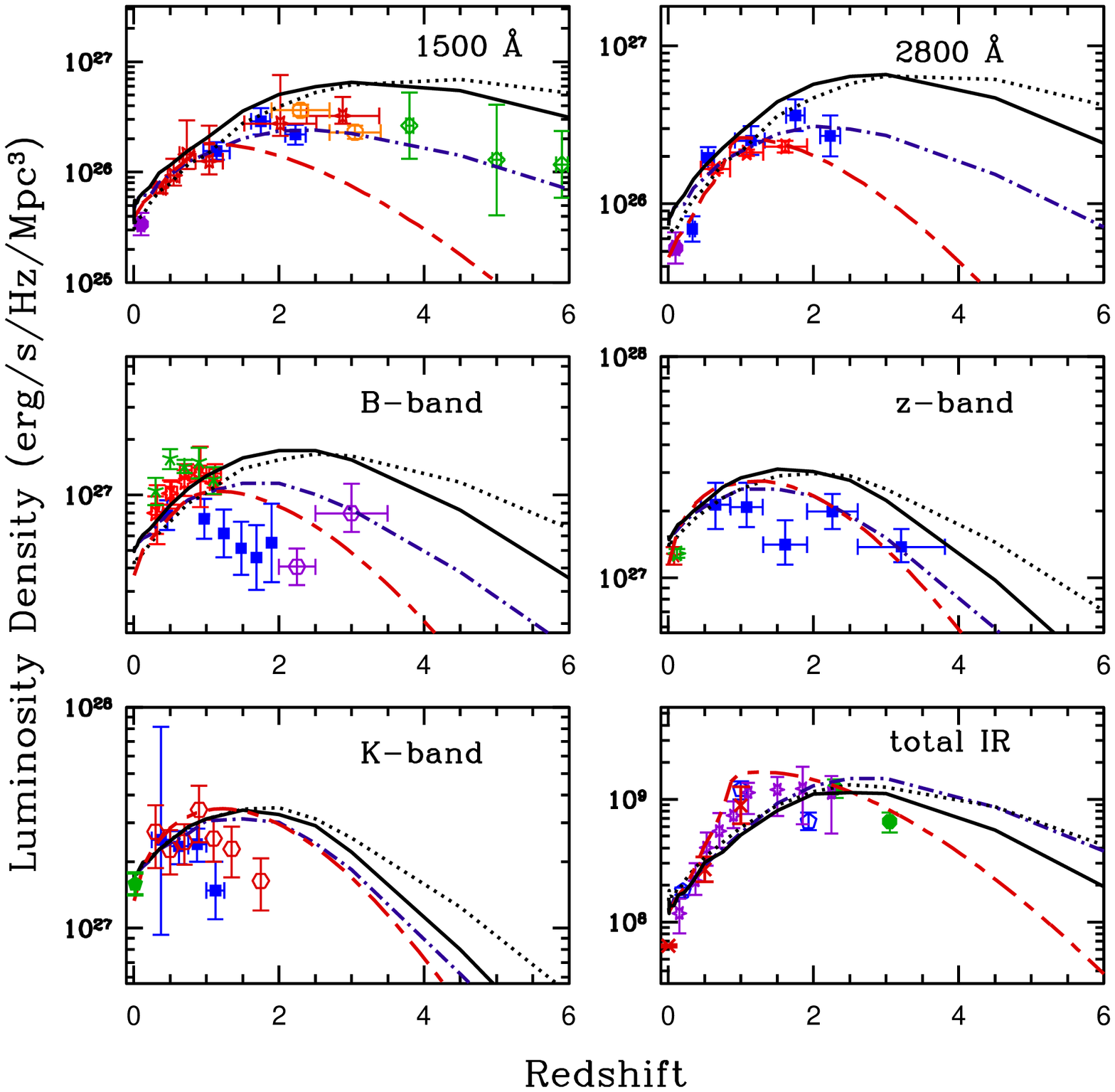,width=15cm}
\caption{The luminosity density (integrated luminosity function of sources within a given redshift range) in our models vs.~redshift at 1500 \AA, 2800 \AA, and in the B-, z-, and K- bands (approximately 4500~\AA, 9130~\AA, and 2.2 $\mu$m, respectively).  The final panel on the lower right shows the total amount of energy that is absorbed and reradiated by dust at infrared energies; units on the y-axis for this panel are solar luminosities per Mpc$^3$.   The solid black line is our WMAP5 prediction with evolving dust, and the dotted line are prior results from our C-\LCDM~model.  Dash-dotted violet shows the predictions from our WMAP5 model using fixed dust attenuation parameters.  The long-short dashed red line is the prediction of D11.  Observational data shown here are as follows: {\bf 1500 \AA:} Blue squares are
from \citet{dahlen07}, red stars are from \citet{schiminovich05},
green stars are from \citet{bouwens07}, and orange circles are from
\citet{reddy08}.  The solid purple circle is a local measurement with
GALEX by \citet{wyder05}. {\bf 2800 \AA:} Blue squares and the purple
circle are again from \citet{dahlen07} and \citet{wyder05},
respectively.  Red stars are from \citet{gabasch06}.  {\bf B-band:}
Blue squares are from \citet{dahlen05}.  Deep and Combo-17 data from
\citet{faber07} are shown as red stars and open red squares,
respectively (these are very similar).  This Combo-17 data is an update to that originally presented in \citet{wolf03}, and we show the original points as green stars.  The work of \citet{marchesini07} is shown as open purple hexes.  {\bf z-band:} Local
measurements are provided by \citet{md&prada09} (red) and
\citet{blanton03} (green).  Blue squares are from \citet{gabasch06}.
      {\bf K-band:} The local determination is from
      \citet{kochanek01}.  High redshift data are from \citet{barro09}
      (blue squares) and \citet{arnouts07} (open red circles).  {\bf
        Total IR Luminosity:} observational estimates of the IR
      emissivity are from \citet{caputi07} (open blue pentagons),
      \citet{reddy08} (green circles), \citet{rodighiero10} (purple
      stars), and \citet{lefloch05} (red crosses).}
\label{fig:ldevo}
\end{figure*}

We have also compared our model with observational estimates of the
evolving luminosity density at a number of different redshifts (Figure
\ref{fig:ldevo}).  The peak emissivity redshift in our model changes
depending on the wavelength considered.  At UV bands, the emission
closely follows the star-formation rate, which peaks at $z\approx 2.5$
in our WMAP5 model and $z\approx 3$ in the C\LCDM~model.  Longer
wavelengths include significant contributions from progressively more
evolved stellar populations, and therefore peak at later times.
Recent evolutionary surveys such as DEEP2 and COMBO-17 allow us to
compare the evolution of galaxy emissivity against accurate luminosity
density data in several bands.  Emissivity in the UV (1500 and 2800
\AA) has been seen to increase out to nearly $z=2$ \citep{dahlen07}.  Note the large discrepancy between the evolving and fixed dust
attenuation models at high redshift in the B-band and UV; in the
evolving model much less attenuation of starlight occurs in early
star-forming galaxies.   At high redshift our fiducial UV predictions are somewhat higher than the measurements of \citet{bouwens07}; this is largely due to a substantial contribution to the total emissivity from faint galaxies which have very little extinction in our evolving dust model.

In the B-band, \citet{dahlen05} find that emission increases out to at
least $z=1$; this paper makes the claim that emissivity in the B- and
R-bands is consistent with being flat in the interval $1<z<2$.
Results at the higher redshifts could be sensitive to the faint end
slope assumed in calculating the luminosity density.  In the K-band we match well the local luminosity measurement of \citet{kochanek01}.  Available data at higher redshifts  seem to suggest a falloff in emissivity beginning at about $z\gtrsim 1$ which we do not find in our models.  As discussed in \citet{sgpd11}, our model does seem to overpredict the K-band luminosity of galaxies at and below L$^*$,
beginning at redshifts 1 to 2.  A corresponding overproduction of NIR
flux is not seen in the local luminosity density or K-band counts.

\subsubsection{Local EBL flux and discrete sources}

As mentioned in the Introduction, measurements of the local ($z=0$) EBL generally fall into two categories: direct sky photometry and integrated counts of galaxies.  Direct measurements provide an absolute measurement of the background light without regard to the sources responsible, but require subtraction of foreground sources present in the Milky Way and our solar system in order to isolate the extragalactic signal.   Integration of galaxy counts (galaxies per unit sky area at a given magnitude) is a way to set firm lower limits on the EBL, although the degree to which these measurements converge on the true value often remains controversial.    The flux from faint sources will converge mathematically if the slope of the counts plotted on a log number vs flux diagram is flatter than unity, or in terms of magnitudes if $\alpha < 0.4$, for $\ln(N) \propto \alpha \, m$.   As expounded by \citet{bernstein07}, photometry of faint galaxies is fraught with difficulty in untangling the faint galactic fringes from the background, and it is possible to miss 50\% or more of the light associated with extended sources in simple aperture photometry.   

Large scale surveys such as the Sloan Digital Sky Survey (SDSS), the 6-degree Field survey (6dF) and the 2-Micron All Sky Survey (2MASS) have provided us with an accurate accounting of the galaxies in the local universe, and surveys with the HST have complemented this data with extremely deep counts.  Satellite instruments such ISOCAM, IRAC, and MIPS provide data in the mid- and far-IR.  A detailed presentation of galaxy number counts in our models compared
with data can be found in SGPD11.  

Our prediction for the local EBL is generally in agreement with lower limits from integrated number counts.  In the UV, limits from \citet{gardner00}, are considerably higher than the the measurement from GALEX \citep{xu05}.   This may be explained by the
former's use of data from the balloon-based FOCA experiment to find
bright counts, which were in disagreement with those from GALEX at
several magnitudes.  Preliminary Herschel counts
data from \citet{berta10} set only a weak lower limit on the FIR
background peak, and the author acknowledges that only about half the
total IR background is likely being resolved.

Absolute measurements of the EBL require the removal of foreground sources, including stars, ISM emission, and sunlight reflected from dust in the inner solar system (often called `zodiacal' light).   
The most robust direct measurements of the IR background to date come from the Diffuse Infrared Background Experiment (DIRBE) and Far-Infrared Absolute Spectrophotometer (FIRAS) instruments on the Cosmic Background Explorer (COBE) satellite, though they are still fraught with uncertainty in sky subtraction (see Figure 2 in \citet{hauser&dwek01}).   The near-IR flux has been calculated from DIRBE observations by a variety of authors \citep{wright&reese00,wright01,gorjian00,cambresy01,levenson07} using foreground source subtraction techniques and modeling of the zodiacal light, and has generally yielded high estimates in this range compared to number counts.  Another notable attempt to measure the near-IR background was \citet{levenson08}, which used IRAC data to calculate the best-fit flux at 3.6 $\mu$m using a profile-fit to estimate the light from the unobservable faint fringes of galaxies.  These results were 70 per cent higher than those of the aperture method of \citet{fazio04}, highlighting the large uncertainties that galaxy fringe issues can bring to EBL measurement.

\begin{figure*}
\psfig{file=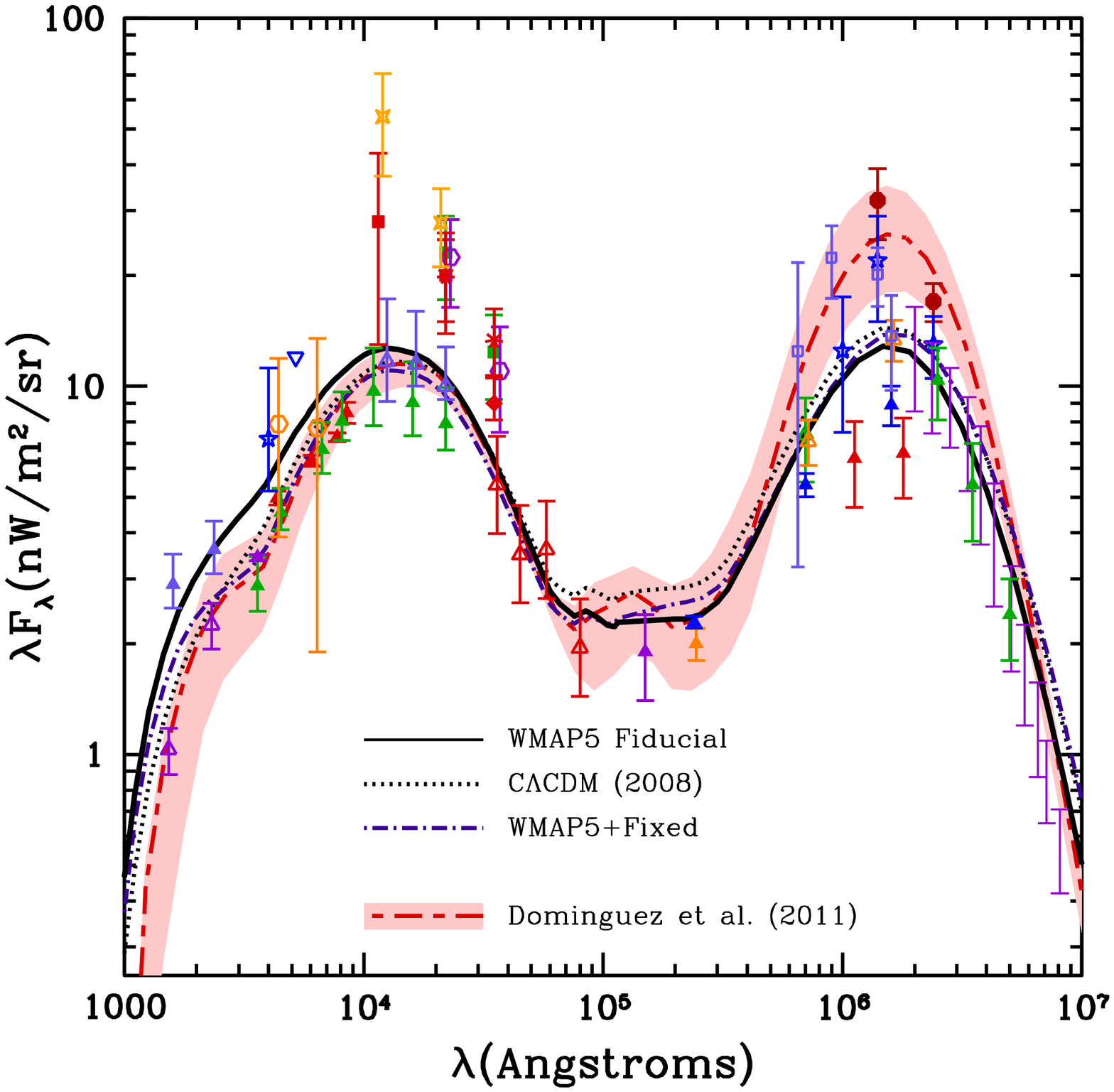,width=15cm,height=13cm}
\caption{ The predicted $z=0$ EBL spectrum from our fiducial WMAP5 model (solid black) and WMAP5+fixed (dash-dotted violet) dust parameters, and C\LCDM~(dotted black) models,  compared with experimental constraints at a number of wavelengths.  
D11 is shown for comparison in dashed-dotted red with the shaded area indicating the uncertainty region.
{\bf Data:} Upward pointing arrows indicate lower bounds from number
counts; other symbols are results from direct detection experiments.
Note that some points have been shifted slightly in wavelength for clarity.  {\bf
  Lower limits:} The blue-violet triangles are results from Hubble and
STIS \citep{gardner00}, while the purple open triangles are from GALEX
\citep{xu05}.  The solid green and red triangles are from the Hubble
Deep Field \citep{madau00} and Ultra Deep Field \citep{dolch11}
respectively, combined with ground based-data, and the solid purple
triangle is from a measurement by the Large Binocular Camera
\citep{grazian09}.  In the near-IR J, H, and K bands, open violet
points are the limits from \citet{keenan10}.  Open red triangles are
from IRAC on Spitzer \citep{fazio04}, and the purple triangle at 15
$\mu$m is from ISOCAM \citep{hopwood10} on ISO.  The lower
limits from MIPS at 24, 70, and 160 $\mu$m on Spitzer are provided by \citet{bethermin10} (solid blue) and by 
\citet{chary04}, \citet{frayer06}, and \citet{dole06} (solid gold, open gold, and open green, respectively).  Lower limits from Herschel number counts \citep{berta10} are shown as solid red triangles.  In the submillimeter, limits are presented from the BLAST experiment (green points; \citealp{devlin09}). {\bf
  Direct Detection:} In the optical, orange hexagons are based on data from the Pioneer 10/11 Imaging Photopolarimeter \citep{matsuoka11}, which are consistent with the older determination of \citet{toller83}.  The blue star is a determination from \citet{mattila11}, and the triangle at 520 nm is an upper limit from the same.  The points at 1.25, 2.2, and 3.5$\mu$m are based
upon DIRBE data with foreground subtraction: \citet[][dark red
  squares]{wright01}, \citet[][orange crosses]{cambresy01},
\citet[][red diamond]{levenson08}, \citet[][purple open
  hexes]{gorjian00}, \citet[][green square]{wright&reese00}, and
\citet[][red asterisks]{levenson07}.  In the far-IR, direct detection
measurements are shown from DIRBE \citep[][blue stars and solid red
  circles]{wright04,schlegel98}, and FIRAS \citep[][purple
  bars]{fixsen98}. Blue-violet open squares are from IR background measurements
with the AKARI satellite \citep{matsuura10}. }
\label{fig:eblflux}
\end{figure*}

The present-day EBL obtained in each of our models is shown in Figure
\ref{fig:eblflux}.  We also show results from D11 for comparison.  The local EBL is calculated by integrating over the luminosity
density at all wavelengths beginning at $z=7.5$, and accounting for
the redshifting and dilution of photons as the universe expands.  The
EBL at a redshift $z_0$ and frequency $\nu_0$ in proper coordinates
can be written as \citep{peebles93}
\begin{equation}
J(\nu_0,z_0)=\frac{1}{4\pi} \int^{\infty}_{z_0} \frac{dl}{dz} \frac{(1+z_0)^3}{(1+z)^3}\;\epsilon (\nu,z)\; dz,
\end{equation}
where $\epsilon(\nu,z)$ is the galaxy emissivity at redshift $z$ and frequency $\nu=\nu_0(1+z)/(1+z_0)$, and $dl/dz$ is the cosmological line element, which is
\begin{equation}
\frac{dl}{dz}=\frac{c}{(1+z)H_0} \frac{1}{\sqrt{\Omega_m(1+z)^3+\Omega_\Lambda}}
\label{eq:cosline}
\end{equation}
for a flat $\Lambda$CDM universe.  We assume here that the EBL photons evolve passively after leaving their source galaxies and are not affected by any further interactions except for cosmological redshifting.  This is an acceptable approximation for photons at energies below the Rydberg energy of 13.61 eV.  Photons above this energy are strongly attenuated by neutral hydrogen when leaving their galaxy of origin.  At higher energies, photons are also capable of interacting with residual neutral hydrogen and, if sufficiently energetic, neutral and singly-ionized helium in the intergalactic medium.  The effect of these processes on the ionizing EBL is the topic of our previous work in \citet{gilmoreUV}.  The total flux of the integrated EBL, as well as the contributions from the optical--near-IR and far-IR peaks and the mid-IR valley for each model are shown in Table \ref{tab:eblintegrated}.  

\begin{table*}
 \begin{tabular}{@{}ccccc}
  \hline
  Wavelength Range & WMAP5 (Fiducial) & WMAP5+Fixed  & C\LCDM  & D11\\
  \hline
Optical--near-IR peak (0.1 to 8 $\mu$m) & 29.01 & 24.34 & 26.15 & 24.47\\
Mid-IR (8 to 50 $\mu$m) & 4.89  & 5.16 & 5.86 & 5.24\\
Far-IR peak (50 to 500 $\mu$m) & 21.01 & 22.94 & 24.08 & 39.48 \\
Total (0.1 to 500 $\mu$m) & 54.91 & 52.44 & 56.09 & 69.19\\
\hline

  \hline
 \end{tabular}
 \caption{The integrated flux of the local EBL in our models (WMAP5 with evolving and fixed dust parameters, and the C\LCDM~model) and the model of D11.  Units are nW/m$^{2}$/sr.}
 \label{tab:eblintegrated}
\end{table*}

\subsubsection{Evolution of the background flux}

A correct determination of gamma-ray opacity at distances beyond the
very nearby universe, $z>0.05$, requires accounting for the
redshift-dependent evolution of the background at all wavelengths.
The sharply increasing star formation rate density from $z=0$ back to
$z\sim2$, combined with the $(z+1)^4$ evolution of proper flux in
redshift means that the background was considerably more powerful in
the recent past, a fact that can only be neglected in gamma-ray
attenuation calculations for the closest extragalactic sources.  With
observations of VHE extragalactic sources now stretching out to
redshifts of over 0.5, it is important in comparing different
realizations of the EBL that we focus not only on the flux at $z=0$
but at higher redshifts as well, where behavior may be quite different
in a given model depending on how galaxies evolve.  We show how the
background develops in our models in two ways in Figure
\ref{fig:eblhist}.  The top panels show the proper EBL SED from
different redshifts in the rest frame, for each of our models. The
bottom panels show the comoving EBL at those same redshifts; this is
the background that would be seen today if all galaxy emissivity had
been shut off below the indicated redshift.  It can be seen in the top
plots that the EBL photon density was considerably higher in the past
at all wavelengths.  The most striking increases from present day
levels are in the mid- and far-IR, and in the UV.

Complementary to Figure \ref{fig:eblhist}, in Figure
\ref{fig:phothist} we show how the photons populating the IR EBL at
various wavelengths today were produced as a function of redshift.  As
expected from our knowledge of obscured starbursting galaxies at high
redshift, the mid- and far-IR parts of the EBL came into existence
considerably sooner than the photons that are part of the
optical--near-IR peak today.  Our results are in reasonable agreement
with a recent survey of submillimeter galaxies \citep{devlin09}, which
has found that half of the background radiation at 250 $\mu$m is
produced at $z>1.2$, with this fraction increasing at longer
wavelengths.  The results for the WMAP5 and C\LCDM~ models are
qualitatively similar, however due to earlier star formation in the
latter, a greater percentage of photons were in place at a given
redshift for all wavebands relative to the WMAP5 model.   {Because the measurements shown here are unavoidably incomplete at high redshift, the fact that our models overpredict the fraction of light in place at early times is not necessarily in conflict with these results.

\begin{figure*}
\psfig{file=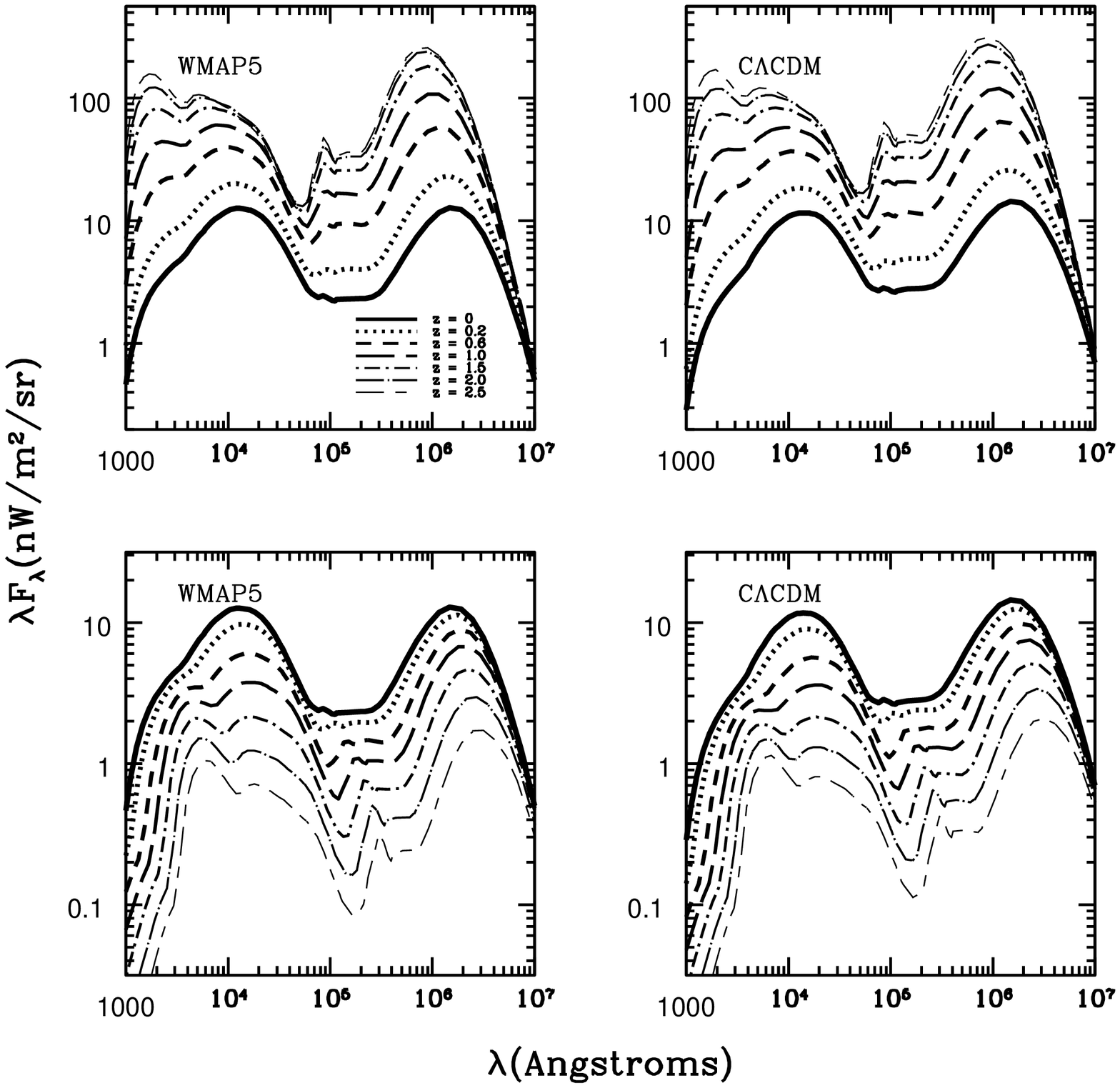,width=15cm}
\caption{The history of the EBL in each of our models.  The top 2 plots show the background flux at past redshifts in the WMAP5 fiducial (left) and C\LCDM~(right) models in standard units.  Redshifts shown include $z=0$ (solid), $z=0.2$ (dotted), $z=0.6$ (short dashed), $z=1$ (long dashed), $z=1.5$ (dot-short dashed), $z=2$ (dot-long dashed) and $z=2.5$ (long and short dashed); also see the key in the upper-left panel.  The bottom two plots show the same quantities, but now evolved to present-day (co-moving), allowing easy comparison of the EBL in place at a particular time compared to the total in existence at $z=0$.}
\label{fig:eblhist}
\end{figure*}

\begin{figure} 
\centering
\psfig{file=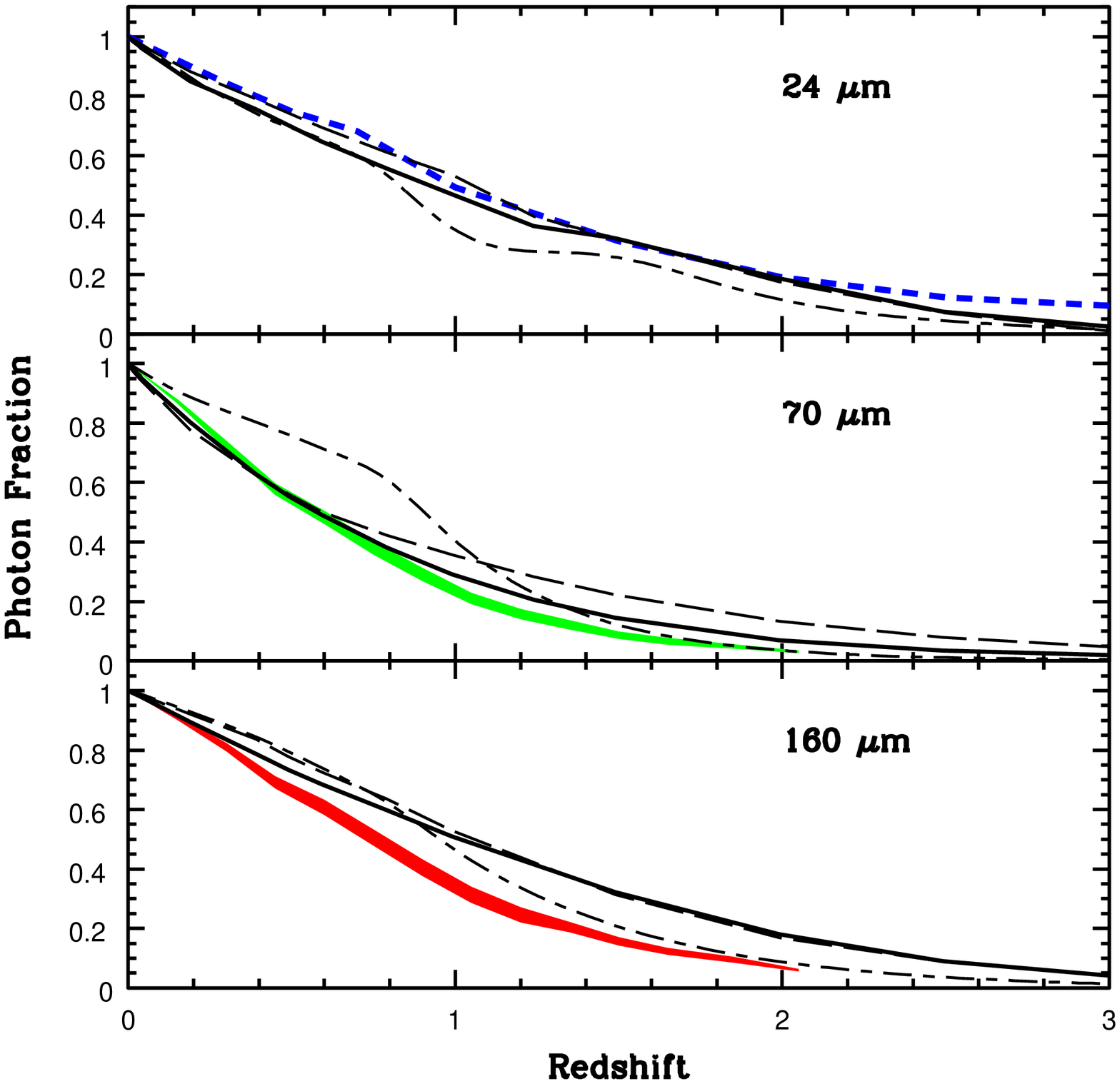,width=7.5cm}

\caption{Buildup of the present-day background photon population in 3  different bands (observed-frame), showing the fraction of the present-day co-moving number density in place at a given redshift.  Solid black is our fiducial WMAP5 model, dashed is the WMAP5+fixed, and dash-dotted is the model of D11.  We have compared against the measurements of \citet{lefloch09} (blue) at 24\mum~and \citet{jauzac11} at 70  (green) and 160 \mum~(red).  For the latter two, the extent of the shaded region is the 1$\sigma$ error bound.
\label{fig:phothist}}
\end{figure}

The rapid increase in flux at all wavelengths with increasing redshift means that the attenuation per unit distance increases a corresponding amount.  Therefore, gamma rays from more distant blazars suffer more attenuation than might be expected from the local EBL flux.  In addition, the SED shape of the EBL changes, so a simple $z$-dependent scaling factor is not sufficient to allow accurate predictions of spectral modification for the more distant sources.

\subsection{Gamma-ray attenuation}
\label{subsec:gammaatt}

\label{gammarayatt}
The process of photon-photon scattering to electron-positron pairs is well understood from quantum electrodynamics.  The basic kinematic requirement for this process is that there must be sufficient energy in the center-of-mass frame of the two-photon system to create the pair.  Including the effect of interaction angle as measured in the cosmological frame, this can be written
\begin{equation}
\sqrt{2E_1 E_2 (1-\cos\theta)} \geq 2 m_e c^2,
\label{eq:paircre}
\end{equation}
where $E_1$ and $E_2$ are the photon energies and $\theta$ is the angle of incidence.  We are interested here in cases where the target background photon has energies from the far-IR ($\gtrsim 10^{-2}$ eV) to the Lyman limit ($\lesssim 13.6$ eV).  The corresponding gamma-ray energies are therefore in the TeV or GeV range.   
We can rewrite Equation \ref{eq:paircre} to define the minimum threshold energy $E_{th}$ for a background photon to interact with a gamma ray of energy $E_{\gamma}$,
\begin{equation}
E_{th}=\frac{2m_e^2c^4}{E_\gamma (1-\cos\theta)}.
\end{equation}
\noindent The cross-section for this process is
\citep{gould&schreder67,madau&phinney96}
\begin{align}
& \sigma(E_1,E_2,\theta) = \frac{3\sigma_T}{16}(1-\beta^2) \times \nonumber \\  & \times \left[ 2\beta(\beta^2-2)+ (3-\beta^4)\ln \left( \frac{1+\beta}{1-\beta}\right)\right], 
\label{eq:sigma}
\end{align}
where
\begin{equation}
\beta = \sqrt{1-\frac{2m_e^2c^4}{E_1 E_2 (1-\cos\theta)}},
\end{equation}
and $\sigma_T$ is the Thompson scattering cross section.

The cross section is maximized for center-of-mass energies of approximately twice the threshold energy $2m_e c^2$, and falls approximately as inverse energy for $E \gg E_{th}$.  If we also account for $\theta$, we find that the likelihood of absorption is maximized for photons at about 4 times the absolute threshold energy, with one factor of 2 from $\sigma$ and another in going from $\theta=\pi$ (`head-on' configuration) to the most probable angle of interaction $\theta \approx \pi /2$.  If we assume $\theta=\pi/2$, then we can define the characteristic energy or wavelength for the background photons that will most strongly affect a gamma ray of energy $E_\gamma$ as
\begin{equation}
E_{bg}=\frac{4 m_e^2 c^4}{E_\gamma} = 1.044 \; \left(\frac{\mbox{TeV}}{E_\gamma}\right) \;\mbox{eV},
\end{equation}
\noindent or equivalently,
\begin{equation}
\lambda_{bg}=1.188 \; \left(\frac{E_\gamma}{\mbox{TeV}}\right) \; \mu \mbox{m}.
\label{eq:charwave}
\end{equation}

Gamma rays at a rest-frame energy above 1 TeV are most attenuated by the near- and mid-IR range of the EBL, while those in the 200 GeV to 1 TeV regime are sensitive to light in the near-IR and optical peak in the EBL SED.  Below 200 GeV it is mainly UV photons that have sufficient energy to cause the pair-production interaction.  Below 19 GeV only background photons with energies above the Lyman limit of 912 \AA~have sufficient energy to interact at any angle in the rest frame, and there is little attenuation due to the paucity of such photons \citep{oh01,gilmoreUV}. 

To calculate the optical depth for a gamma ray observed at energy $E_\gamma$, we perform the integral along the line of sight to the target at redshift z,  
\begin{align}
\tau(E_\gamma,z_0) =  \frac{1}{2}\int^{z_0}_0 dz\;\frac{dl}{dz}\int^1_{-1}du \; (1-u) \times \nonumber \\ \times \int^{\infty}_{E_{min}} dE_{bg}\; n(E_{bg},z)\;\sigma(E_\gamma (1+z),E_{bg},\theta).
\label{eq:opdep}
\end{align}
\noindent Where we have \[E_{min}=E_{th}\:(1+z)^{-1}=\frac{2m_e^2c^4}{E_\gamma (1+z)(1-\cos\theta)}\] to account for the redshifting of the gamma-ray energy.  Here $n(E_{bg},z)$ is the proper density of target background photons as a function of energy $E_{bg}$ and redshift $z$, and $u$ is shorthand for $\cos\theta$.  $dl/dz$ is the cosmological line element, Equation \ref{eq:cosline}.

For nearby sources, $z \lesssim 0.05$, it is sufficient to use the local EBL density $n(E_{bg},z=0)$.  However, as we saw in Figure \ref{fig:eblhist}, both the total power and SED of the EBL vary strongly with redshift, and in general it is therefore necessary to understand the evolution of the background to correctly compute gamma-ray opacities.  The rapid increase in flux at all wavelengths with increasing redshift to $z \gtrsim 2$ means that the attenuation per unit proper distance increases a corresponding amount.  This effect means that gammas from more distant blazars suffer more attenuation than might be expected from the local EBL flux.  In addition, the functional form of the EBL changes, so a simple z-dependent scaling factor is not sufficient to allow accurate predictions of spectral modification for the more distant sources.  Using the line-of-sight integral, Equation \ref{eq:opdep}, we show in Figure \ref{fig:opdep} the optical depth vs. gamma-ray energy for a variety of redshifts.  A more general way to show EBL attenuation is to plot the `attenuation edge' redshift where the optical depth reaches a certain value as a function of gamma-ray energy, and this is presented out to high redshift for 3 values of $\tau$ in Figure \ref{fig:attedge}.  This shows how telescopes with lower energy thresholds will allow us to peer deeper into the universe.  See \citet{gilmoreUV} for a similar plot extending to lower gamma-ray energies and higher redshift.

\begin{figure}
\centering
\psfig{file=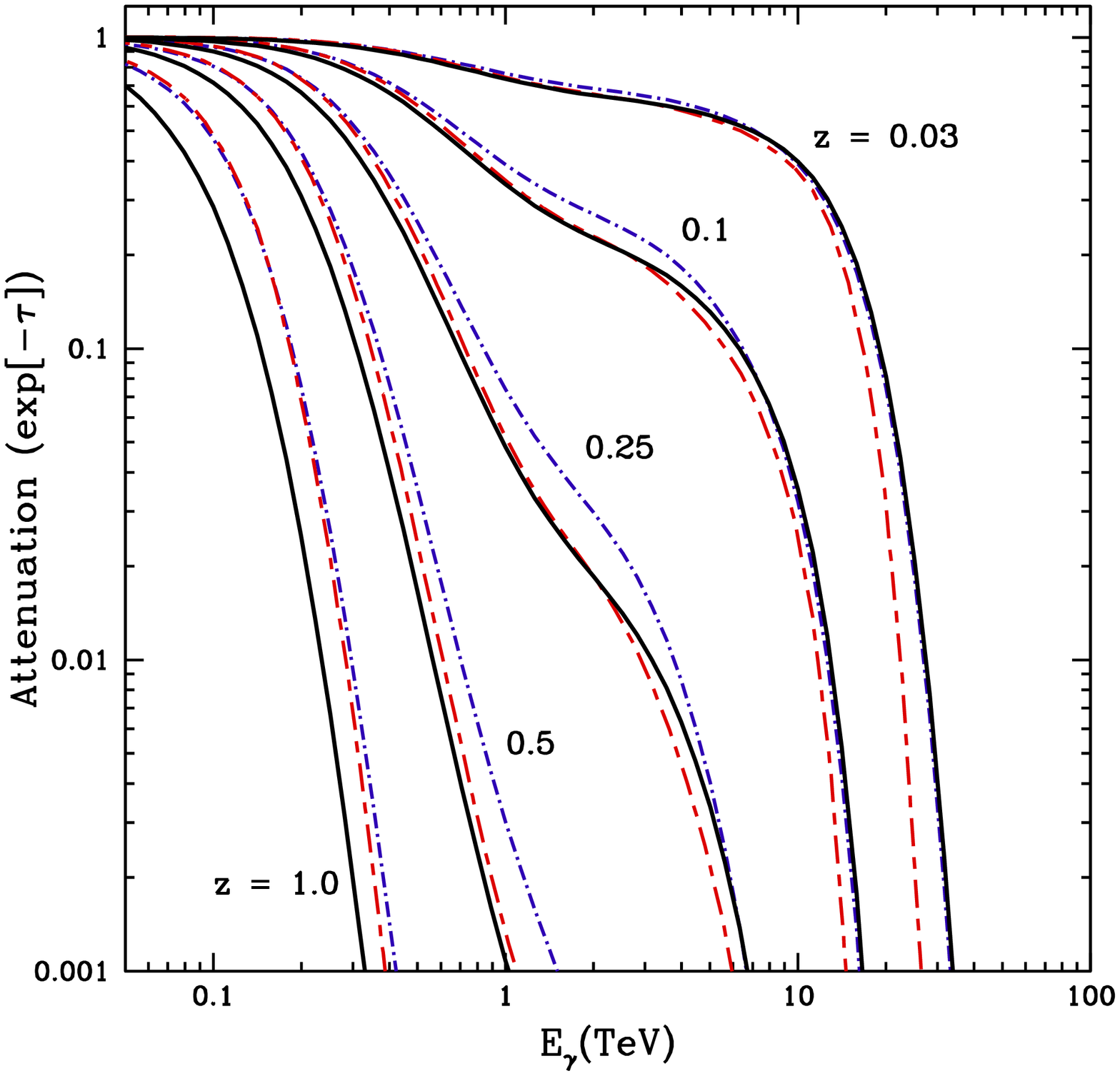,width=8cm}
\caption{The attenuation $e^{-\tau}$ of gamma-rays vs. gamma-ray energy, for sources at $z= 0.03$, 0.1, 0.25, and 0.5, and 1.  Results are compared for our fiducial WMAP5 (solid) and WMAP5+fixed (dash-dotted violet) models, as well as the model of D11 (red dash-dotted).  Increasing distance causes absorption features to increase in magnitude and appear at lower energies.  The plateau seen between 1 and 10 TeV at low redshift is a product of the mid-IR valley in the EBL spectrum.}
\label{fig:opdep}
\end{figure}

\begin{figure}
\centering
\psfig{file=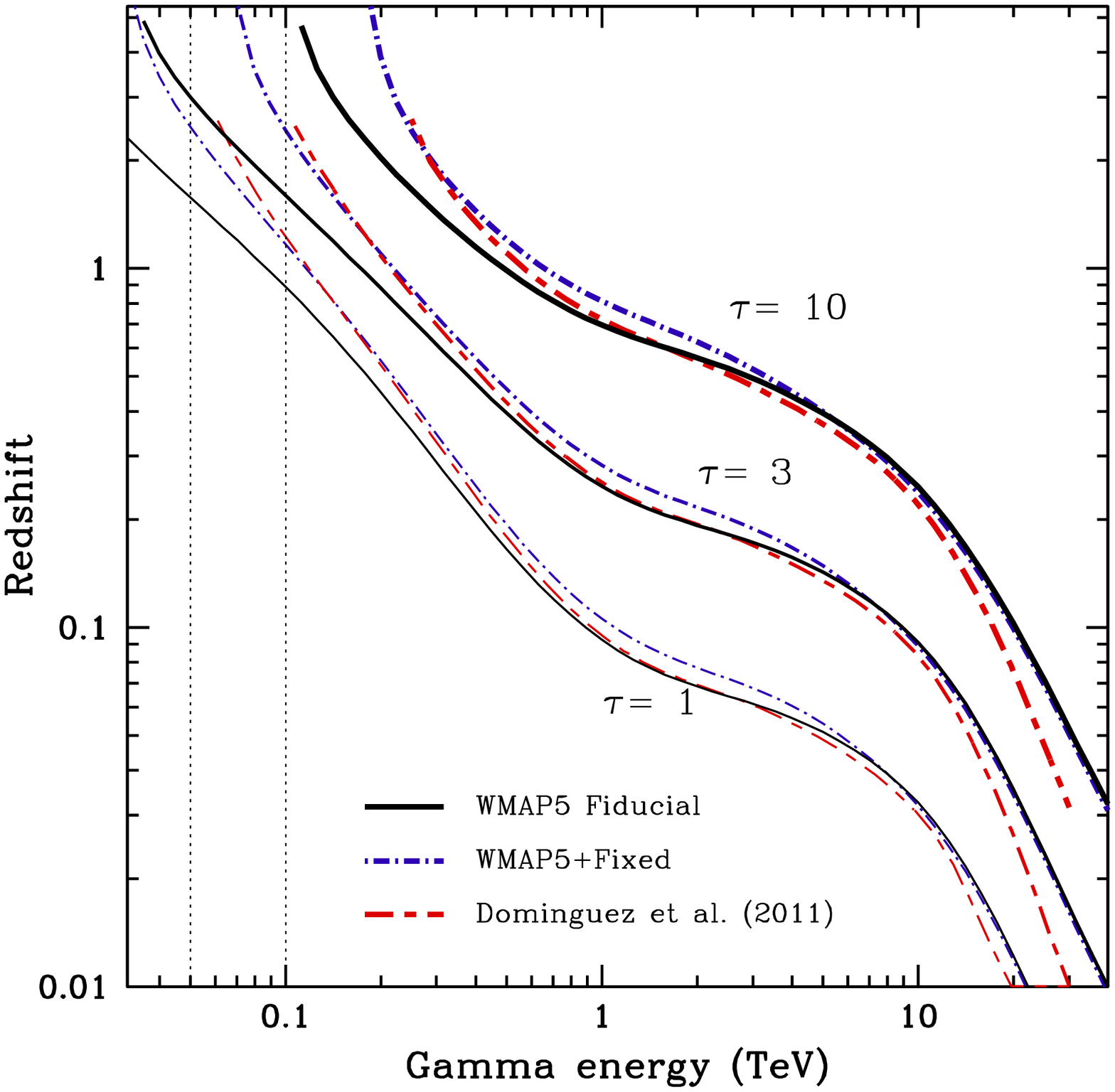,width=8cm}
\caption{The gamma ray attenuation edges for the WMAP5 (solid black) and WMAP5+fixed (dash-dotted violet) models and model of D11 (red dash-dotted).  The curves show the redshift at which the pair-production optical depth $\tau$ reaches the indicated value for a particular observed gamma ray energy.  The groups of curves from lower left to upper right are the contours for $\tau=$ 1, 3, and 10.  We have included thin lines to guide the eye at 50 and 100 GeV.}
\label{fig:attedge}
\end{figure}

\subsection{Results for TeV blazars}
\label{subsec:blazars}
Today, exploration in the VHE (30 GeV to 30 TeV) regime is led by $>$10m-class imaging atmospheric Cherenkov telescopes (IACTs) including the  VERITAS \citep{maier07}, H.E.S.S. \citep{hinton04}, and MAGIC \citep{cortina05} experiments, and by the LAT instrument on the Fermi gamma-ray space telescope \citep{atwood09} and also AGILE \citep{tavani08}.   

The Fermi LAT spends most of its time in an-all sky survey mode, and with its large area of view is therefore an ideal instrument for finding high-energy sources.  The 11-month source catalog lists 685 high-energy sources associated with blazar candidates \citep{abdo10}.   While the Fermi LAT has an energy range of 20 MeV to $\sim$300 GeV, it has a much smaller effective area than the current generation of ground-based instruments,  and data from the instrument is therefore most useful for our purposes at energies below the threshold of these IACTs, 50-100 GeV.  A detailed analysis of the EBL constraints available from all Fermi observations of blazars and GRBs to date was the subject of a recent paper by the Fermi collaboration, \citet{fermiEBL}.  Current limits on the EBL available from Fermi observations do not constrain the UV flux predicted in \citet{gilmoreUV} or in the models presented here.  

In this section and the following section, we will focus on the effect of the optical-IR EBL on AGN-type sources by IACTs at $\gtrsim$100 GeV.  Ground-based detectors searching above 100 GeV have identified 37 extragalactic AGN-like sources at the time of this writing, including 32 BL Lac objects, radio galaxies M87 and Centaurus A, and the flat-spectrum radio quasars 3C279, PKS 1510-08, and PKS 1222+21.  With the exception of the radio galaxies these objects are all blazars, accreting AGN which generate tightly beamed relativistic jets that are oriented at a small angle relative to our line of sight.  While they account for the large majority of detected sources above 100 GeV, BL Lac objects are themselves only a small subset ($\sim$20$\%$) of all blazar sources, the other 80 percent being flat spectrum radio quasars like 3C279.

\subsubsection{Constraints from gamma-ray observations}

While uncertainties and likely variation in the intrinsic spectrum of blazars make it impossible to directly link the observed spectrum to EBL attenuation, it is possible to translate limits on the spectra to EBL constraints.  The standard assumption in placing limits on the EBL from individual spectra is that the reconstructed intrinsic spectrum should not have a spectral index harder than 1.5; that is, $\Gamma \geq 1.5$ where $dN/dE \propto E^{-\Gamma}$ for photon count $N$, or alternatively $dF/dE \propto E^{-(\Gamma-1)}$ for flux $F$.  This figure comes about both on the basis of experimental observations (no observed VHE spectrum is harder than this value) and theoretical arguments.  The standard value for a single-zone synchrotron-self Compton (SSC) spectrum is $\Gamma=(\alpha+1)/2$; here $-\alpha$ is the spectral index of the shock-accelerated electrons, which is not harder than 2.0 in most acceleration models with radiative cooling \citep{aharonian01}. 
This can be invalidated by assuming a non-standard spectrum for the electrons; a low energy cutoff in the electron energy will lead to inverse-Compton accelerated photons with an index as low as $\Gamma= 2/3$ \citep{katarzynski06}.  

The most recent limits on the EBL come from observations of blazars at more distant redshifts ($z>0.1$) that have been detected by the current generation of ground-based atmospheric Cherenkov telescopes (ACTs).  Observation by H.E.S.S. of two blazars at z=0.165 and 0.186 were used to set limits on the near-IR EBL based on the $\Gamma \geq 1.5$ criterion \citep{aharonian06}; in this case the maximal limit was the model of \citet{primack01} multiplied by a factor of 0.45.  Another paper by the H.E.S.S. group set constraints from blazar 1ES 0229+200 at z=0.1396 \citep{aharonian07}.  While this blazar is a closer source than the two featured in the 2006 publication, the observed spectrum extended above 10 TeV and therefore probed the background in the mid-IR (Equation \ref{eq:charwave}).  In this regime, the effect of optical depth on spectral modification is minimal due to the approximate $\lambda^{-1}$ falloff in EBL flux, which leads to a constant photon density per logarithmic bin, and therefore an approximately constant gamma-ray opacity as a function of energy.   The most distant source observed at very high energies at the time of writing is quasar 3C279 at z=0.536, observed by the MAGIC experiment during a flare in February 2006 \citep{teshima08}.  The spectrum observed was quite steep, $4.1\pm0.7_{stat}\pm0.2_{sys}$, and extended from about 80 to nearly 500 GeV.  An analysis of the spectral modification \citep{albert08} found that there was little room for an EBL flux in the optical higher than one consistent with lower limits from number counts, approximately equivalent to the model of \citet{primack05}.  This paper used a modified version of the `best fit' model from \citet{kneiske04} as the upper limit in the optical and near-IR from their finding.  An alternative analysis of the spectral deconvolution of 3C279 by \citet{stecker&scully09} disputed this analysis and argued that the higher EBL of \citet{stecker06,stecker07} could still lead to a steep best-fit spectrum.  However, this higher EBL is inconsistent with Fermi high-redshift blazar observations at the $5\sigma$ level \citep{fermiEBL}.

Another approach to the problem is to attempt to constrain the EBL by using spectra from several sources simultaneously.  \citet{dwek&krennrich05} derived an upper limit at 60$\mu$m by declaring invalid those realizations leading to unphysical intrinsic blazar spectra with sharply rising TeV emissions.   More recently, this method was used in \citet{mazin&raue07}, who applied constraints from all observed TeV blazars to a large number of possible EBL functional forms created using a spline interpolation across a grid in flux versus wavelength space.  The lower bound of the union of excluded models formed an envelope representing the highest possible background that does not violate any constraints.  This was done for `realistic' and `extreme' bounds of $\Gamma \geq 1.5$ and $2/3$ respectively, and provided a limit on the EBL from the optical to the far-IR. The latter case is motivated by the limiting case of a truncation at a low energy bound for the relativistic electrons responsible for the IC component, see \citet{katarzynski06}.

In Figure \ref{fig:eblflux_gamlims}, we show recent upper limits from gamma-ray observations in relation to the $z=0$ EBL from our models.  All of our models are generally in agreement with these bounds across all wavelengths.   While our fiducial model does slightly exceed the bounds set by \citet{albert08} in the optical, we do not find a conflict with the standard $\Gamma \geq 1.5$ bound in Table \ref{tab:blazars}, and we are within $1\sigma$ agreement with the harder spectrum observed by \citet{aleksic11}.

It is worth pointing out here that, in general, one should use caution concerning these constraints.  These limits on the present-day EBL do not take into account the differences in redshift evolution occurring in different EBL models, which becomes increasingly problematic for more distant sources.  Also, as mentioned above the limits from \citet{aharonian06} and \citet{albert08} assume specific forms for the optical peak of the background SED with variable normalization.  The exact normalization of the upper bound is dependent upon this choice.  The method used by Mazin \& Raue avoids this second issue, but at a cost of more conservative limits resulting from considering a finite grid in flux--wavelength space. 

\begin{figure}
\centering
\psfig{file=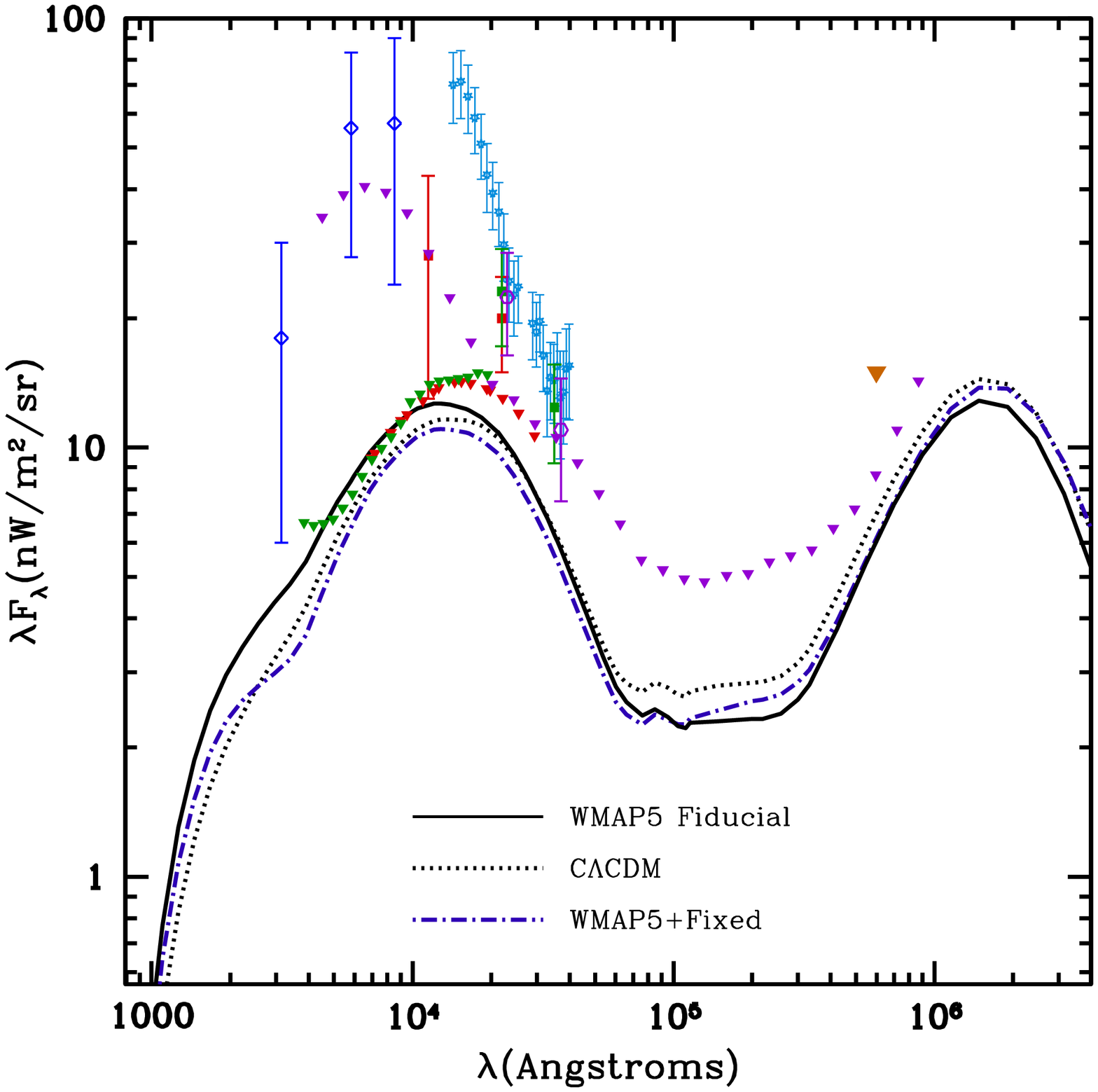,width=8cm}
\caption{Present-day flux predicted in our EBL models, compared against upper limits from gamma-ray observations.  Line types are the same as in Figure \ref{fig:eblflux}.  Upper limits are shown as downward-pointing triangles.  Red triangles are observations described in \citet{aharonian06}, and green are from MAGIC observations of 3C279 \citep{albert08}.  Other limits  shown are the realistic limits of \citet{mazin&raue07} (purple), and the analysis of \citet{dwek&krennrich05} (orange triangle at 60$\mu$m).  The reader should consult the text for more details and caveats in interpreting this figure.  We also show for comparison many of the high estimates of the optical and near-IR EBL from direct photometry.  The open blue diamonds are from \citet{bernstein07}.  The points at 1.25, 2.2, and 3.5$\mu$m are based upon DIRBE data with foreground subtraction: \citet{wright01} (dark red squares), \citet{gorjian00} (purple open hexes), and \citet{wright&reese00} (green square).  The small cyan points are from direct photometry with the IRTS satellite \citep{matsumoto05}. }
\label{fig:eblflux_gamlims}
\end{figure}

The photon density of the EBL increases with wavelength at almost all energies relevant to gamma-ray attenuation, and therefore the effect on high energy spectra is always a spectral softening.  However, it is possible that local radiation in the vicinity of a source could have other effects on the spectrum.  This is particularly true for FSRQ sources such as 3C279.  As a quasar, 3C279 is a much more powerful source at optical and UV wavelengths than BL Lac objects.  It has therefore been suggested that internal absorption from the broad-line region of the quasar could harden the spectrum by creating an optical depth that decreases with energy over the observed interval \citep{aharonian08}, due to emission in a narrow band of UV wavelengths.  An analysis by \citet{tavecchio&mazin09} claimed that while significant internal absorption was likely, only the more extreme models of the broad line region lead to an actual hardening of the intrinsic spectrum, and these models lead to a large decrease in flux from absorption, by a typical factor of $>10^3$.  This effect could potentially harden intrinsic spectra emerging from AGN beyond the bounds discussed above, but only in limited extreme cases.  More reasonable models of the local radiation fields with less total absorption were found to leave the spectral index softened or unmodified. 

\subsubsection{Spectral modification of known TeV blazars}
\label{sec:smktb}

We have calculated absorption from each of our EBL realizations in the observed spectra of known blazars that are approximated by power-law functions, and determined the approximate power law of the de-absorbed spectra.  The spectra from these objects are not expected to be power laws over large energy ranges.  The most simple theoretical form of the spectra from SSC emission is a double-peaked distribution (when plotted as $\nu F_{\nu}$), which arises from synchrotron radiation of lower energy photons and inverse Compton upscattering of those same photons to gamma-rays.  In this model, the power law measured at VHE scales is an approximation to a section of the inverse Compton peak.  
	
Also, the effect of gamma-ray attenuation through pair production does not in general preserve a power-law form, as can be seen in the optical depth plot, Figure \ref{fig:opdep}.  Quantifying attenuation as a simple modification to an intrinsic spectral index is an approximation that is only valid when considering short intervals in energy and fairly low redshifts.  The EBL attenuation has also been described as an decaying exponential function in energy that affects the spectra above some threshold.  However, this is a misleading functional description of the optical depth.  The sharp increase in absorption in Figure \ref{fig:opdep} that appears at multi-TeV energies is caused by the rapid increase in photon density as one transitions from the mid-IR minimum in the EBL SED and into the redshift-broadened PAH region and far-IR peak (note that our SED is plotted in terms of flux density, not number density).  This part of the EBL is created by re-emitted light from cold dust, much of which originates in rapidly star-forming galaxies, and there is no reason to believe that this absorption feature would be related to an exponential form.  The power law and the exponential cut-off, which are often used to describe gamma-ray spectra, are not amenable to describing the full non-linear effects of EBL absorption, which is a line-of-sight integral over the evolving photon field.  Our optical depths for nearby sources are relatively straight from a couple hundred GeV out to this turnover region, so we present results for sources with spectra measured in this energy range.  

One other note concerns the integration over bins of finite width in energy.  As attenuation differs across these intervals, it changes the weighting of data and therefore the mean within the bin.  Properly de-absorbing spectral data points requires incorporating the optical depth into the analysis used to produce the points, and not just multiplying by $e^\tau$ at the mean of the bin.  Correlations between the data points must also be accounted for in effectively measuring error.  The effect of scaling the data with a simple multiplication introduces error that is likely to grow with redshift.

\begin{table*}
\centering
\scriptsize
\begin{tabular}{@{}lcllcccccc}
  \hline
  Object ID & Redshift & Reference & Experiment & $E_{L \gamma}$  & $E_{H \gamma}$ & $\Gamma_{obs}$ & $\Gamma_{\mbox{\tiny Fiducial}}$ & $\Gamma_{\mbox{\tiny Fixed}}$ \\
\hline
Mrk 421 (+) & 0.030 & \citet{konopelko08} & Whipple & 0.2 & 8.0 & 2.66 & 2.46 & 2.46  \\
\hline
Mrk 421 & 0.030 & \citet{aharonian99} & HEGRA & 0.5 & 7.0 & 3.09 $\pm$ 0.07 & 2.88 & 2.88\\
\hline
1ES 2344+514 & 0.044 & \citet{albert07} & MAGIC & 0.14 & 5.4 & 2.95 $\pm$ 0.12 $\pm$ 0.20 & 2.71 & 2.72\\
\hline 
Mrk 180 (+) & 0.045 & \citet{albert06b} & MAGIC & 0.2 & 1.5 & 3.30 $\pm$ 0.70 & 3.04 & 3.06\\ 
\hline
1ES 1959+650 & 0.047 & \citet{albert06} & MAGIC & 0.18 & 2.0 & 2.72 $\pm$ 0.14 & 2.46 & 2.48\\
\hline
1ES 1959+650 (-) & 0.047 & \citet{tagliaferri08} & MAGIC & 0.15 & 3.0 & 2.58 $\pm$ 0.18 & 2.34 & 2.36\\
\hline
1ES 1959+650 (+) & 0.047 & \citet{daniel05} & Whipple & 0.38 & 18.0 & 2.78 $\pm$ 0.12 $\pm$ 0.21 & 1.95 & 1.91 \\
\hline
BL Lacertae & 0.069 & \citet{albert07a} & MAGIC & 0.15 & 0.9 & 3.60 $\pm$ 0.50 & 3.25 & 3.29\\ 
\hline
PKS 0548-322 & 0.069 & \citet{superina08} & H.E.S.S. & 0.2 & 3.0 & 2.8 $\pm$ 0.33 & 2.41 & 2.45 \\
\hline
PKS 2005-489 & 0.071 & \citet{aharonian05a} & H.E.S.S. & 0.2 & 2.5 & 4.0 $\pm$ 0.4 & 3.60 & 3.63\\ 
\hline
RGB J0152+017 & 0.080 & \citet{aharonian08c}  & H.E.S.S. & 0.24 & 3.8 & 2.95 $\pm$ 0.41 & 2.48 & 2.51\\
\hline
W Comae (+) & 0.102 &\citet{cogan08} & VERITAS & 0.15 & 2.8 & 3.81 $\pm$ 0.35 $\pm$ 0.34 & 3.26 & 3.32 \\
\hline
PKS 2155-304 & 0.116 & \citet{aharonian05} & H.E.S.S. & 0.16 & 0.70 & 3.32 $\pm$ 0.06 & 2.74 & 2.81 \\
\hline
H 1426+428 & 0.129 & \citet{aharonian02} & HEGRA & 1.0 & 6.0 & 2.60 $\pm$ 0.60 $\pm$ 0.1 & 1.74 & 1.67 \\
\hline
1ES 0806+524 & 0.138 & \citet{acciari09} & VERITAS & 0.3 & 0.7 & 3.6 $\pm$ 1.0 $\pm$ 0.3 & 2.68 & 2.79 \\ 
\hline
1ES 0229+200 & 0.139 & \citet{aharonian07} & H.E.S.S. & 0.5 & 7.0 & 2.50 $\pm$ 0.19 $\pm$ 0.10 & 1.45 & 1.40\\
\hline
H 2356-309 & 0.165 & \citet{aharonian06} & H.E.S.S. & 0.16 & 1.0 & 3.06 $\pm$ 0.40 & 2.11 & 2.23\\ 
\hline
1ES 1218+304 & 0.182 & \citet{albert06a} & MAGIC & 0.09 & 0.63 & 3.00 $\pm$ 0.4 & 2.28 & 2.39\\  
\hline 
1ES 1218+304 & 0.182 & \citet{fortin08} & VERITAS & 0.16 & 1.8 & 3.08 $\pm$ 0.34 $\pm$ 0.20 & 2.02 & 2.14 \\
\hline
1ES 1101-232 & 0.186 & \citet{aharonian06} & H.E.S.S. & 0.16 & 3.3 & 2.88 $\pm$ 0.17 & 1.83 & 1.91 \\ 
\hline
1ES 0347-121 & 0.188 & \citet{aharonian07a} & H.E.S.S. & 0.25 & 3.0 & 3.10 $\pm$ 0.23 $\pm$ 0.10 & 1.94 & 2.03 \\ %
\hline
1ES 1011+496 (+) & 0.212 & \citet{albert07b} & MAGIC & 0.12 & 0.75 & 4.00 $\pm$ 0.50 & 2.95 & 3.10 \\ %
\hline
S5 0716+714 (+) & 0.31$\dagger$ & \citet{mazin09} & MAGIC & 0.2 & 0.7& 3.45 $\pm$ 0.54 & 1.47 & 1.74\\
\hline
PKS 1222+21 (+) & 0.432 & \citet{aleksic11a} & MAGIC & 0.07 & 0.4 & 3.75 $\pm$ 0.27 $\pm$ 0.2 &  2.32 & 2.60 \\
\hline
3C66A (+) & 0.44$\dagger$ & \citet{acciari09a} & VERITAS & 0.2 & 0.5 & 4.1 $\pm$ 0.4 $\pm$ 0.6 & 1.34 & 1.75 \\
\hline
3C279 (+) & 0.536 & \citet{albert08} & MAGIC & 0.09 & 0.48 & 4.11 $\pm$ 0.68 & 2.71 & 2.14 \\ %
\hline
3C279 (+) & 0.536 & \citet{aleksic11} & MAGIC & 0.15 & 0.35 & 3.1 $\pm$ 1.1 & 0.51 & 0.97 \\ %
\hline

 \end{tabular}

\medskip
 \caption{Reconstruction of the VHE spectral indices of a number of blazars using our two WMAP5 EBL realizations.  $\Gamma_{obs}$ is the index reported by the given reference at energies between $E_{L \gamma}$  and $E_{H \gamma}$, reported in TeV.  These are taken from the reference if explicitly stated, otherwise the highest and lowest data points presented are used.  In some cases the highest energy data point presented has large error bars or does not match well with the power-law fit, and we have opted to use the second highest point instead.  $\Gamma_{\mbox{\tiny Fiducial}}$ and $\Gamma_{\mbox{\tiny Fixed}}$ are the average intrinsic indices after de-absorption by our two EBL models, over the range of energies claimed in the detections.  Errors on this quantity are the same as in the observed indices, if provided by the author.  Plus (+) and minus (-) after the source name are used to signify that the detection was claimed in an abnormally high or low state; readers should consult the references given for further details.  Many of these references were taken from \citet{wagner08} and also the TeV online catalog, {\it http://tevcat.uchicago.edu/}. 
 \label{tab:blazars}
\newline
\medskip
\newline
$\dagger$ The redshifts of 3C66A and S5 0716+714 quoted here are considered uncertain; see references for details.}
\end{table*}

Having warned the reader of these caveats, we present results for known blazars seen above $\sim 100$ GeV in Table \ref{tab:blazars}.  The results in the table are also shown in graphical form in Figure \ref{fig:blazars}.  This plot shows the amount of change in spectral index after EBL deconvolution for a number of blazars, as a function of source redshift.  The majority of the objects presented here are of the high frequency-peaked BL Lac (HBL) type, with the exceptions of intermediate-peaked W Comae and 3C66A, low-peaked BL Lacertae and S5 0716+714, and flat spectrum radio quasars 3C279 and PKS 1222+21.  Values from this table should only be taken as approximate, particularly for more distant sources.   To avoid the hazards of analyzing data bin-by-bin, as mentioned in the previous paragraph, we have based our results here on the spectral indices and applicable energy ranges as derived by the authors referenced, whenever possible.  In the cases where the spectrum is claimed to continue above the turnover in optical depths seen at several TeV, the results become strongly dependent upon the highest energy extent of the fit.  When an explicit value is not mentioned, we use the highest energy point displayed in the spectrum in the reference.   It is argued in at least a couple of cases (Markarian 421 and 501, \citep{konopelko03}) that the de-absorbed spectrum shows the rollover at the top of the IC peak.  Recent MAGIC observations of Mrk 501 have detected a spectral peak at energies which vary in correlation with flaring activity \citep{albert07c}.  As simple power-law functions do not provide a good fit in this case, we have omitted Mrk 501 from our analysis.  

Two spectra in our analysis show unusually hard reconstructed spectra.  In the case of H 1426+428 \citep{aharonian02}, the reported spectrum that we have used does not conform well to a power law, a fact that the authors attribute to EBL absorption, and to be cautious we have used an upper energy of 6 TeV.  The spectral index of 1ES 0229+200 \citep{aharonian07} is sensitive to the highest energies used in the calculation, and we assume here an upper energy of 7 TeV, which is close to the second-highest data point presented in the reference.    In general, the reconstructed spectral indices for spectra extending above a few TeV are highly sensitive to the highest energy data point considered due to the rapid increase in opacity with energies at this scale (Figure \ref{fig:opdep}).

It is also interesting to compare the effect that dust modeling has on spectral reconstruction, by comparing the results of the fixed and evolving (fiducial) treatments in the WMAP5 model.  The former produces an EBL that is slightly more intense in the mid-IR, while at shorter wavelengths the evolving dust treatment of the fiducial model allows a higher optical and near-IR intensity.  Higher redshift blazars have generally only been seen at lower energies, where gamma-ray attenuation is produced by the optical-near IR EBL peak created mostly by redshifted direct starlight.  As can be seen in Table \ref{tab:blazars} and Figure \ref{fig:blazars}, the larger optical and near-IR flux of the fiducial model produces more spectral change than the fixed model in most blazars.  For a few blazars that have been seen to multi-TeV energies, the fixed model produces a harder slope.  

\begin{figure}
\centering
\psfig{file=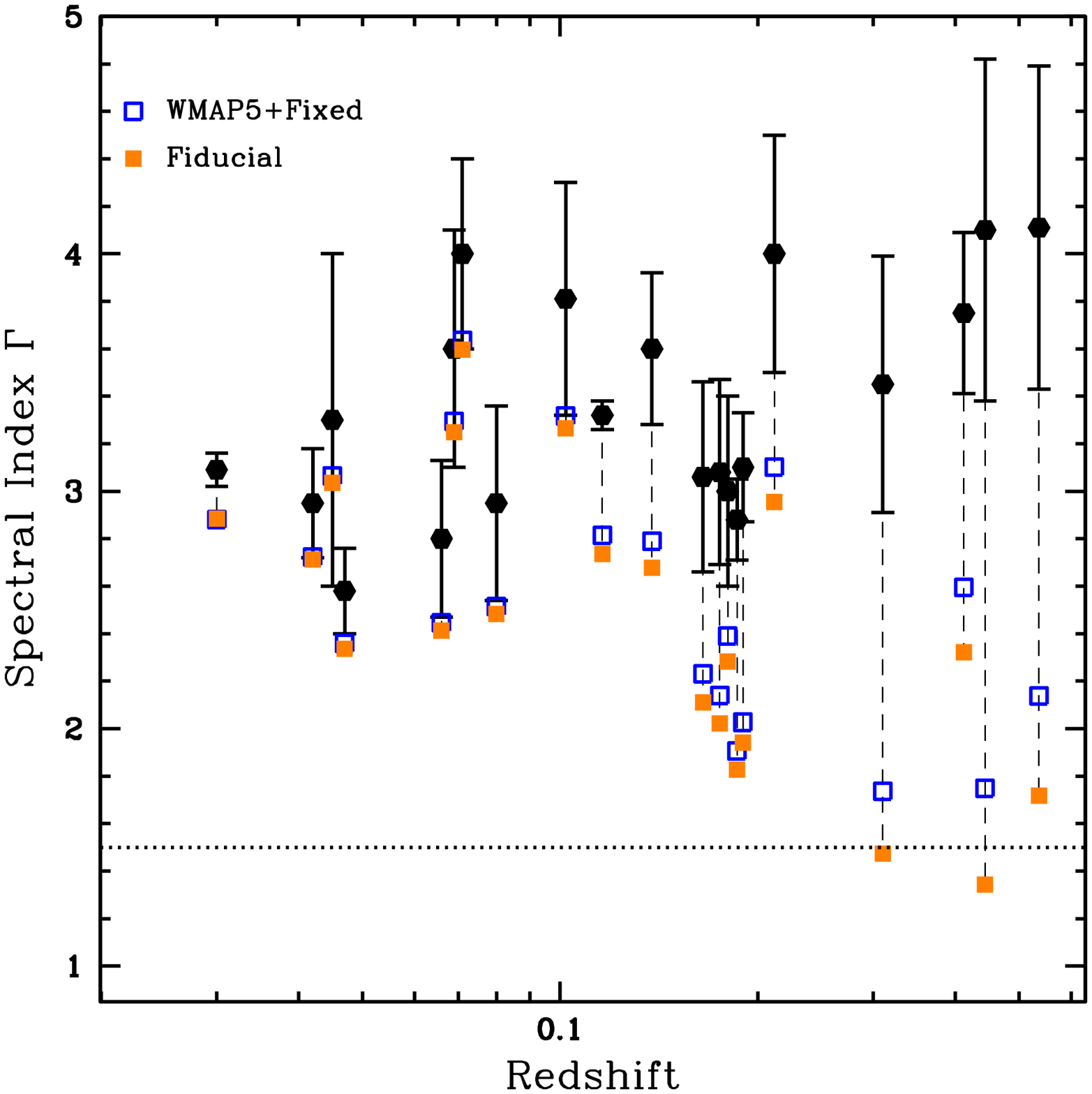,width=8cm}
\caption{Here we show the results of Table \ref{tab:blazars} in graphical form.  The measured spectral index ($\Gamma$; $dN/dE \propto E^{-\Gamma}$) and redshift of each blazar is shown as a black hexagon with error bars, with the index corrected via the Fiducial EBL shown as a solid orange point, and that corrected by the WMAP5+Fixed model as an open blue point.  The horizontal dotted line shows $\Gamma=1.5$, which is typically taken as the hardest spectrum possible under usual assumptions.  Some points have been shifted sideways slightly for readability. }
\label{fig:blazars}
\end{figure}

As mentioned in \citet{acciari09a}, determining the redshift of blazars can be difficult due to the lack of strong line emission.   Blazars S5 0716+714, PG 1553+113, and 3C66A are cases where VHE spectra have been published, but the source redshift remains uncertain.  With a given background model, gamma-ray attenuation can be used to place upper redshift limits on these sources (e.g., \citealp{prandini10,yang&wang10}).  Applying the standard $\Gamma \geq 1.5$ constraint, we have summarized some findings for these redshift constraints in Table \ref{tab:blazespectz}.  Here, we use the upper $1\sigma$ bound on source spectral index as the basis for the limit.  The redshift of 3C 66A was initially determined to be 0.444 from a single emission line, taken to be magnesium-II, and corroborated by a weak Ly$\alpha$ detection.  Assuming this is correct, we find for this blazar a reconstructed spectrum that is harder than many others on the list, but still significantly softer than the standard limit at the $1\sigma$ uncertainty bound.  The HBL PG 1553+113 has been detected by both MAGIC \citep{albert07e} and H.E.S.S. \citep{aharonian08a}, but the redshift remains unknown at this time.  Observations with the HST have been unable to find a precise distance, but suggest a redshift in the range $0.3 < z < 0.4$ \citep{treves07}, potentially making this the most distant VHE-detected HBL.  Studies of the intervening Ly$\alpha$ forest with COS \citep{danforth10} have more recently found a higher redshift bound, $z >  0.395$.  Our most stringent constraints with the WMAP5 model put this blazar at $z<0.40$ with the spectral index presented by \citet{prandini09}, or $z<0.48$ if the index is taken at the upper $1\sigma$ level.  A similar analysis by \citet{mazin&goebel07} using MAGIC data and a low-level EBL similar to the level set by galaxy counts found $z<0.69$, while \citet{prandini09} placed a limit $z<0.67$ with the model of \citet{kneiske&dole10}.  This blazar was also considered in D11, and an upper redshift of $z \leq 0.85 \pm 0.07$ was found.  Reconstructing the intrinsic spectrum at higher redshifts also was found to lead to a break in the power-law shape.  Demanding that such a break be absent leads to a tighter upper limit in this reference, $z < 0.42$.  The weight of this evidence would seem to suggest a redshift near the lower limit set by Danforth.  S5 0716+714 is an LBL for which a spectrum has been recently reported by MAGIC, and was previously detected by EGRET and AGILE.  Our fiducial bound of $z<0.31$ ($z<0.37$ at $1\sigma$) is in agreement with the range reported by \citet{nilsson08}, $z=0.31\pm0.08$.

\begin{table*}
\centering
\begin{tabular}{@{}lcllcccccc}

  \hline
  Object ID & Reference & Experiment & $E_{L \gamma}$  & $E_{H \gamma}$ & $\Gamma_{obs}$ & z & z \\
  & & & & & & {\tiny Fiducial} & {\tiny Fixed} \\
\hline
S5 0716+714 (+)  & \citet{mazin09} & MAGIC & 0.2 & 0.7 & 3.45 $\pm$ 0.54 & $<$0.37 & $<$0.42 \\
\hline
PG 1553+113 (+)  & \citet{wagner08a} & MAGIC & 0.09 & 0.5 & 4.2 $\pm$ 0.3 & $<$0.62 & $<$0.72 \\
\hline
PG 1553+113 & \citet{prandini09} & MAGIC & 0.15 & 0.6 & 4.12 $\pm$ 0.17 & $<$0.48 & $<$0.51 \\
\hline
PG 1553+113 & \citet{aharonian08a} & H.E.S.S. & 0.23 & 1.3 & 4.5 $\pm$ 0.32 & $<$0.48 & $<$0.52 \\
\hline
3C66A (+) & \citet{acciari09a} & VERITAS & 0.2 & 0.5 & 4.1 $\pm$ 0.4 $\pm$ 0.6 & $<$0.53 & $<$0.56 \\
\hline
\end{tabular}

\medskip
 \caption{Here we show the upper redshift limits for 3 blazars with uncertain redshift based on the $\Gamma \geq 1.5$ criterion discussed in the last section, and using the upper $1\sigma$ uncertainty bound on the spectral indices.  Some of these blazars were also shown in Table \ref{tab:blazars}, assuming there the most likely source redshift.}
\label{tab:blazespectz}
\end{table*}

\section{Comparison with other work}
\label{sec:comparison}

In this section we compare the methodology and results of our EBL
determination with others in the recent literature, including the
previous predictions of earlier versions of our semi-analytic model.
Our prior prediction for the EBL, presented in \citet{primack05}, used
a similar semi-analytic model of structure formation to that which we
have used in this work and in \citet{primack08}.  The WMAP5 model
presented here has similar normalization to the 2005 model in the
optical and near-IR, leading to low flux in these wavebands, with only
a small amount of light unresolved in the deepest number count
surveys.  The differences in the spectral shape of the optical peaks
are due to changes in the application of the dust absorption
prescriptions; in this work and in \citet{primack08} we use the
two-component model of \citet{charlot&fall00}, which leads to more
absorption in the UV and emission in the mid- and far-IR.  A
comparison of the different emission templates was presented in
SGPD11.  The C\LCDM~model we have presented features a higher level
of star-formation, particularly at early times, as a result of
assuming a larger normalization in the initial dark matter power
spectrum.  In the mid- and far-IR, all of our new models produce
significantly more light due to a larger energy budget from absorbed
starlight.  While the prior prediction was not able to match the level
of light suggested by the number counts with ISOCAM at 15$\mu$m or the
level of far-IR flux inferred from DIRBE and FIRAS, all of our new
models are consistent with constraints from number counts in the mid
and far infrared and with FIRAS, and are near the lower determinations
of the DIRBE instrument at 140 and 240 $\mu$m.  The AKARI measurements
of \citet{matsuura10} and some DIRBE measurements \citep{schlegel98}
do suggest a larger flux in the far-IR peak.

The EBL model of \citet{franceschini08} is based on luminosity functions from a variety of survey data.  Recognizing the need for separate treatments of evolution in different wavelength regimes, this model treats optical and IR components separately, using the recent body of data from Spitzer and other experiments.  Near-IR luminosity function data up to $z=1.4$ is used for the spiral and spheroidal populations, while only local luminosity functions are used for the irregular population. Other local IR data are used to constrain other
regions of the spectrum.  The EBL model presented by these authors is
similar to ours at most wavelengths except the far-IR peak.  As their
model has been derived from the same body of cosmological data that
our own has been compared against, it is not surprising to see similar
predictions at low redshift.  In Figure \ref{fig:eblflux_comp} we show
the EBL evolution and gamma-ray attenuation predictions from our models compared with those of \citet{franceschini08},
as well as D11, model `C' in \citet{finke10}, both predictions of
\citet{stecker06} and the best fit model in \citet{kneiske04}, which
we will discuss in the following paragraphs.  Our
model is seen to evolve similarly to the Franceschini model out to
redshift 1.

\begin{figure*}
\psfig{file=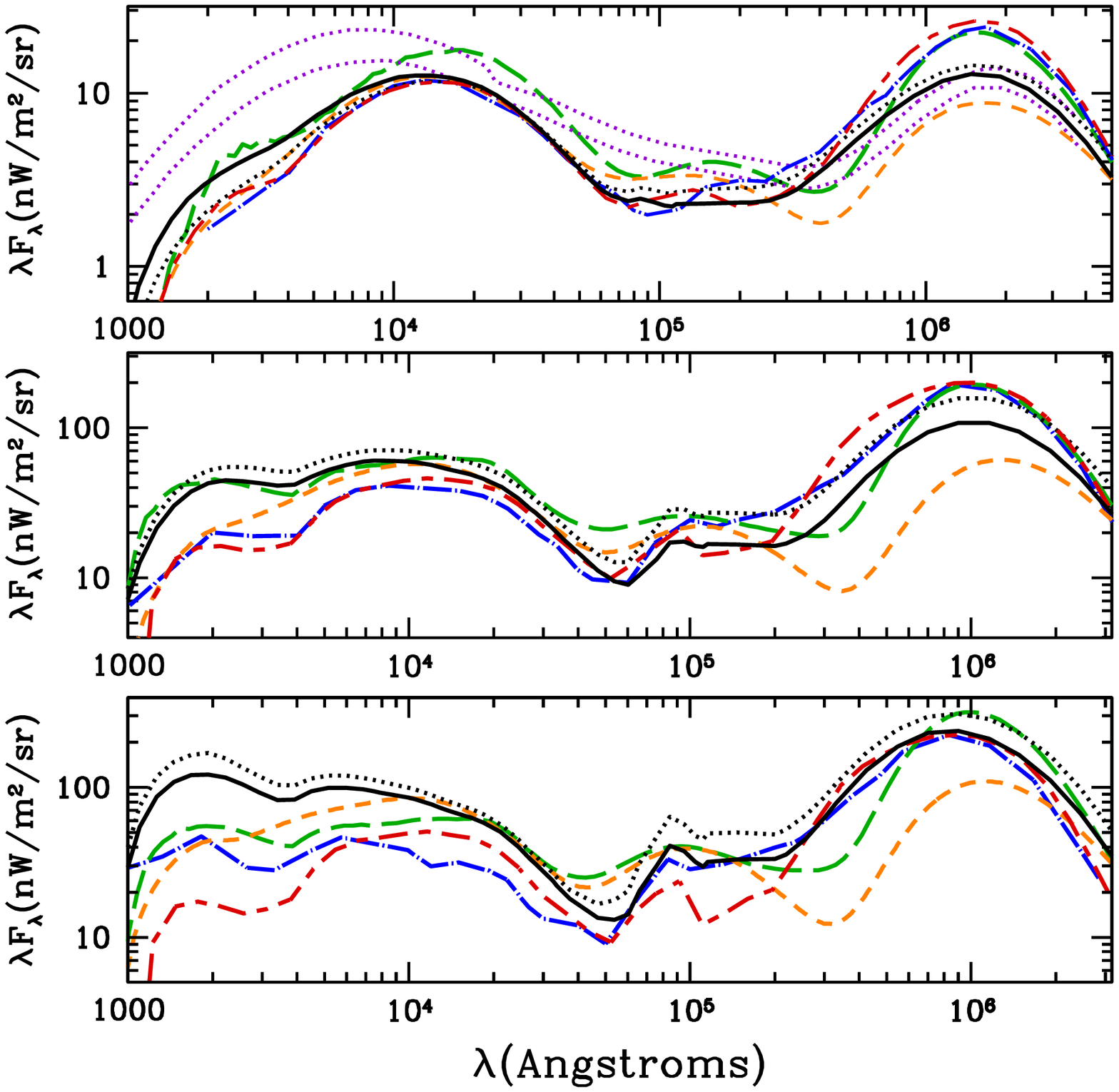,width=8.0cm}
\psfig{file=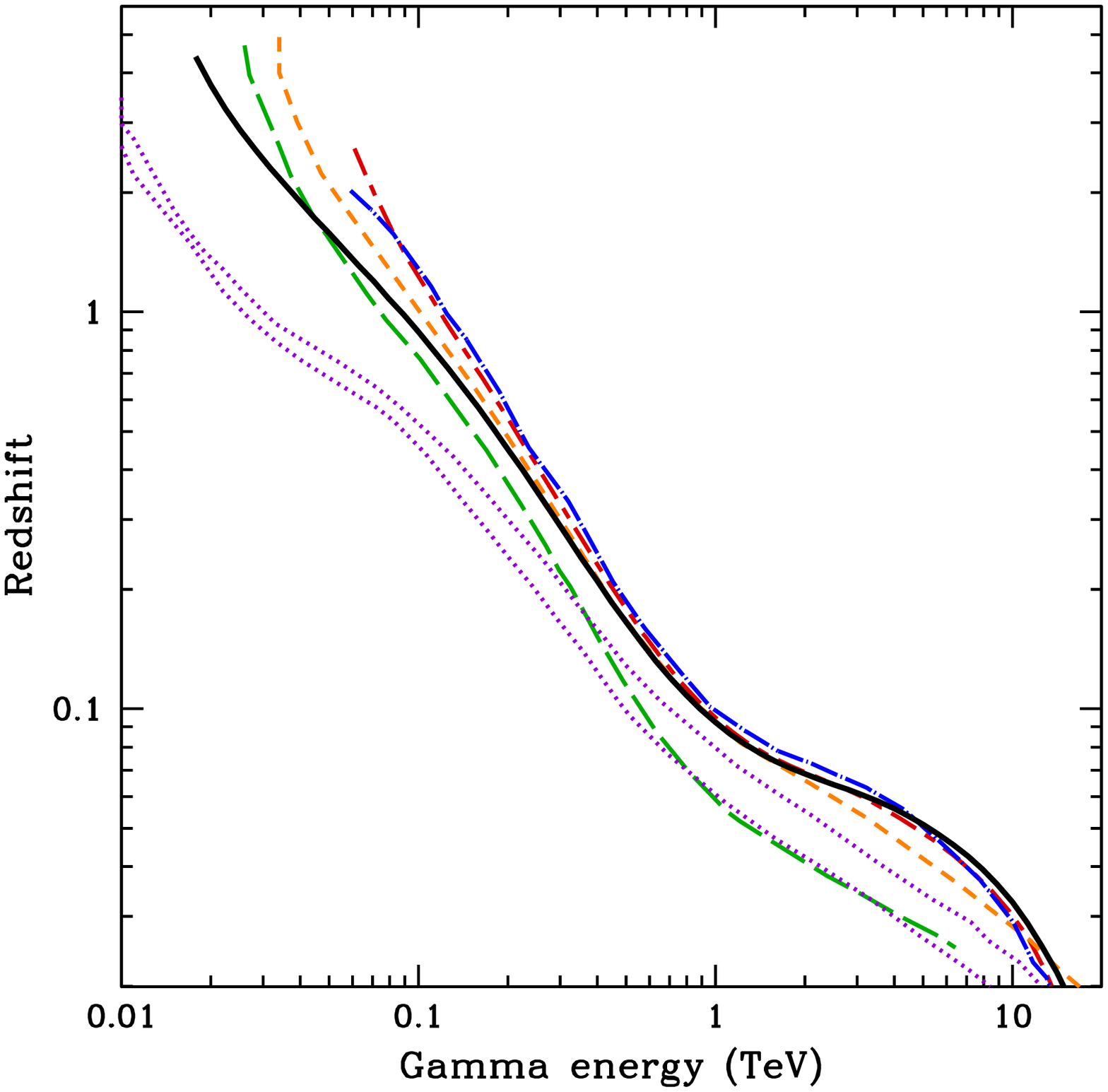,width=8.0cm}
\caption{{\bf Left:} Our EBL predictions compared with several recent models from the literature.  Solid and dotted black lines show the proper flux density from our WMAP5 and C\LCDM~models in the local universe and at $z=1$ and $z=2$.  Other lines are from \citet{franceschini08} (dashed-dotted blue), the best-fit model of \citep{kneiske04} (long-dashed green), D11 (long-short dashed red), and model `C' from \citet{finke10} (dashed orange).  The baseline and fast evolution models of \citet{stecker06} are the low and high dotted violet points in the $z=0$ panel.   {\bf Right:} A comparison of the $\tau_{\gamma \gamma} =  1$ attenuation edges for several models.  Line types are as in the opposite panel.  In this instance, the lower set of dotted points denotes the fast evolution model of \citet{stecker06}, while the upper are the baseline model.}
\label{fig:eblflux_comp}
\end{figure*}

The work of \citet{kneiske02} calculated the EBL from the UV to far-IR
using a `semi-empirical' method based on measured star formation rates
and spectral synthesis models.  Starlight is processed by dust, which
is modeled as a blackbody with three temperature components.
Metallicity is assumed to increase slowly over cosmic time and an
average global extinction curve applied to starlight.  A follow-up
paper, \citet{kneiske04}, expanded this earlier model into 6
realizations, varying in gas temperature contribution, star formation
rate, and UV escape fraction.  The `best-fit' EBL in this paper is
considerably higher than any of our models in both the optical and
far-IR peaks.  The discrepancy in EBL normalization between this model
and our own likely originates in the star formation rate densities
assumed, which have a much different functional form than our model.
Knieske's results are based upon a broken power-law for for the
star-formation history, with a peak at z=1.2.  The history predicted
by our model is considerably lower in this epoch, and does not peak
until z $\approx$ 3 for our C\LCDM~model, or z $\approx$ 2.5 for the
WMAP5 model (Figure \ref{fig:sfr}).  Thus our models have a lower
present-day flux, but higher flux at the location of Kneiske's peak
and at higher redshifts.  The use of a blackbody spectrum to
approximate emission in the PAH region also gives their EBL SED a
somewhat different shape in the mid-IR than we find with our templates
which include these sharp emission features.  The update to this model
in \citet{kneiske&dole10} produces a background flux that is close to
the level seen in discrete number counts from the optical to mid-IR,
and is similar to our Fiducial model.

The recent models presented in \citet{finke10} are based on a similar
technique to Kneiske's work.  These models are built from the earlier
models in \citet{razzaque09}, in which the contribution to the EBL
from main-sequence stars was found by computing stellar emissivities
after assuming forms for the global star-formation history and IMF.
\citet{finke10} expanded these models by including dust re-absorption
and emission, as well as post-main sequence stars.  The authors favor
their model (`C') with a \citet{baldry&glazebrook03} IMF and
star-formation history from \citet{hopkins&beacom06} with
\citet{cole01} parametrization.  The Baldry-Glazebrook IMF produces
slightly more high-mass stars than the Chabrier IMF that we have used.
This model has a slightly higher normalization than ours in the
optical and near-IR.  In the mid- and far-IR, we find considerably
different SED shapes due to the use of different techniques in
modeling dust emission, as was the case with the models of Kneiske et
al.; we have used templates which are based on the galaxies' total IR
emission, while these models assume thermal blackbody emission at
three fixed temperatures.

The models of Stecker and collaborators, most recently
\citet{stecker06,stecker07}, have explored the background using
backward evolution models.  This most recent work proposed two SEDs
for the EBL, using two different assumptions about the pure luminosity
evolution of the 60 $\mu$m luminosity function.  The SEDs of all
galaxies are assumed to be determined by this 60 $\mu$m emissivity.
The `baseline' model features a pure luminosity evolution multiplier
of $(1+z)^{3.1}$ out to z=1.4, and constant luminosity from there to
z=6.  The `fast evolution' model evolves even more quickly, as
$\sim(1+z)^4$ to z=0.8 and $\sim(1+z)^2$ for $0.8<z<1.5$.  Both of
these models are considerably higher than ours in the optical and
near-IR, with the fast evolution model about 50\% higher in this range
than the baseline; the discrepancy in the far-IR with our models is
smaller.  It is difficult to compare our model, which deals with
galaxies in a system of hierarchically merging dark matter halos, with
this model, in which it is assumed that the local galaxy population
just grows brighter with increasing redshift.  Our 60 $\mu$m
luminosity density is not found to increase nearly as quickly as
assumed in either of these models; we find that both of our models can
be well-described by a luminosity density multiplier of
$\sim(1+z)^{1.7}$ out to $z\approx 1.4$ at this wavelength.  As
mentioned in the introduction, the high optical and near-IR flux of
the fast-evolution model puts it at odds with the detection of 3C279
by MAGIC \citep{albert08}, which was disputed by Stecker in another
analysis \citep{stecker&scully09}.  However, the large error on the
determined de-absorbed spectral index ($0.5\pm1.2$), and the
possibility of hardening of the spectrum by internal absorption
(\citealp{liu08,aharonian08}, but see also \citealp{tavecchio&mazin09}),
make it difficult to claim this observation as a firm limit on the
EBL.  Further observations of this and other high-redshift sources
will likely improve constraints on flux in the optical EBL peak.  As
mentioned above, both models in this work are in serious conflict with
the limits set by high-redshift blazars observed by Fermi LAT
\citep{fermiEBL}.

A comparison of the model of D11 with our fiducial model has been
presented and discussed in that work.  Overall, we find good agreement
with D11 from optical to mid-IR wavelengths from the local universe to
$z=1$.  At longer wavelengths, D11 predicts considerably more flux
from starlight reprocessed by cold dust in galaxies.  The large
discrepancy in results for the FIR peak in the EBL highlights the
uncertainties involved in modeling this region, an ongoing challenge
which we address at the end of the next section.  Discrepancies in
this region only have an observable effect on the gamma-ray spectra
for the closest blazars.  The star-formation rate implied in D11 (see
Figure 12 in that work) is considerably higher than our own, and
increases more rapidly with redshift between $z=0$ and $z=1$.  At UV
wavelengths, our model produces more light than predicted in D11, and
more absorption of gamma-rays below $\sim 300$ GeV.  As shown in
SGPD11, the UV luminosity in our model is near the largest values
measured across redshift in integrated luminosity functions, and can
be considered a maximal prediction at these short wavelengths.

At nearly all wavelengths we have considered, our fiducial EBL SED is
near the level of flux resolved in discrete background counts, and we
find agreement with the claim of \citet{madau00} that nearly all light
in the optical EBL peak is produced by discrete sources.  Referring
back to Figure \ref{fig:eblflux}, we recognize two places in our
calculated EBL SED in which there is tension with observations that do
not rely strongly on foreground estimates, and which may signal
shortcomings in our spectral modeling.  In the UV, we find an EBL
lower than calculated using a combination of HDF and balloon-based
FOCA data \citep{gardner00}.  The later GALEX experiment, while not
capable of surveying to the depth of Hubble, found a smaller number of
bright counts than the FOCA data, likely resulting from differences in
calibration of the instruments \citep{xu05}.  It is therefore possible
that the higher Gardner points resulted from overestimating the bright
counts in their determination.  Our prediction for the local EBL is
above the integrated number count measurements of \citet{grazian09} in
the U-band and \citet{dolch11} in the F435W band; in the former there
is significant disagreement with \citet{madau00} at faint magnitudes,
but confirmation that the counts are convergent.  In the near-IR, we find good agreement with published results from \citet{madau00} and \citet{keenan10}.  Our models
fall about 1$\sigma$ below the 5.6 $\mu$m lower bound from Spitzer.
The fact that the 5.6 $\mu$m limit is higher than that at 4.5 may cast
some suspicion on this particular measurement, as there is no reason
to believe such a spectral feature would exist.  These lower limits
are based upon early `first-look' data, and newer results based on a
larger set of survey data should soon be available (G.~Fazio, private
communication).  Additional sensitivity and survey width may be
achievable in the 3.6 and 4.5 $\mu$m bands by post-cryogenic `Warm
Spitzer' surveys; however the longer wavelength bands will not be
attainable at elevated temperatures \citep{vandokkum07}.

Our models lie below the level of direct detection of the absolute background by calculations using data from Hubble WFPC2, DIRBE, and IRTS (see the discussion for Figure \ref{fig:eblflux}), however we do find consistency with the newer optical results of \citet{mattila11}, and also the Pioneer analysis of \citet{matsuoka11}.  The low significance and large error bars on the HST points of \citet{bernstein07} mean that these results should not be considered inconsistent with an EBL at the level provided by resolved sources.  All of our models are at least 1$\sigma$ below the flux from any of the near-IR direct detection calculations we have discussed.  Limits from gamma-ray observations shown in Figure \ref{fig:eblflux_gamlims} have strongly disfavored the highest levels at this range.  As discussed in \citet{levenson07}, the present uncertainty in zodiacal light subtraction may be intractable without a new mission to directly study this foreground, such as the proposed ZEBRA experiment (see discussion in \citealp{zemcov11}).  While our models are consistent within 1$\sigma$ with number count measurements by MIPS in the mid- and far-IR, they are low compared to the DIRBE measurements of the far-IR peak.  The zodiacal foreground is a sharply decreasing function of wavelength in this regime, and the DIRBE points are expected to suffer from less systematic error here than in the near-IR, especially at 240 $\mu$m, where our models lies beneath the data.   However, our IR templates include a somewhat simplified treatment of blackbody emission at these long wavelengths, and we do not expect accurate reproduction in this regime.

\citet{fardal07} compared the possible range of EBL flux measurements with observations of the fossil mass and star formation rate history of the universe.  As our semi-analytic model reproduces these 3 observables, it is worth discussing our work in the context of their claim that a top-heavy or `paunchy' IMF can best fit simultaneously these data.  This proposal is based on the argument that the low levels of estimated stellar mass are difficult to reconcile with the present-day EBL flux suggested as being intermediate between the direct detections and the lower limits from resolved galaxies, which is now known to be in conflict with blazar limits (see Figure \ref{fig:eblflux_gamlims}).  \citet{fardal07} created 3 models of the EBL based on all available observational limits.  Their minimal model, with total flux of 50 nW/m$^{2}$/sr, is set by resolved number counts and is similar to our WMAP5 and C\LCDM~models in the optical and near-IR out to the K-band.  Their best-fit model, based on a compromise between number counts compilations and the HST and DIRBE direct detection measurements, is substantially higher than our C\LCDM~model.  The K-band number counts are well measured by a number of surveys (see Figure 14 in SGPD11) which constrain the amount of stellar mass in the nearby universe.  Two factors alleviate the discrepancy in our model.  Our background fluxes are near the lowest levels considered in \citet{fardal07}, with total fluxes of 56.09 and 54.91 nW/m$^{2}$/sr for our C\LCDM~and WMAP5 models respectively (Table \ref{tab:eblintegrated}), and our global Chabrier IMF produces more high-mass stars than the diet-Salpeter considered as the standard by these authors.  For a near-IR flux much higher than our models to not overproduce the K-band counts, this flux would have to arise from a high-redshift population of sources unresolvable in our current surveys, which extend to $>$ mag 24.  As mentioned in the Introduction, there are reasons why the star-formation rate measures we compare against at high-redshift could be biased high.  This interpretation favors our fiducial EBL model, which has slightly less flux than Fardal's lowest model, and is in fairly good agreement with integrated star-formation and observed K-counts (see SGPD11).

\section{Discussion}
\label{sec:ebldisc}

The EBL presents one of the primary barriers to extragalactic
gamma-ray astronomy with ground-based instruments.  Our determination
of a fairly low extragalactic background across the optical and near
to mid-IR, supported by convergence with alternative methods such as
\citet{franceschini08} and D11 as well as new direct measurement techniques and recent limits from gamma-ray
experiments in the optical to mid-IR, is an optimistic prediction for
the future of the field.  The weight of this evidence also
increasingly points to an EBL that is well-known over this wavelength
range, at a level near that of resolved number counts, though many
questions remain about its redshift evolution.  The ability of
current- and next-generation experiments to detect blazars at larger
distances is a function of several factors: the luminosity function
and spectral evolution of these objects, the effective area
(especially at the lowest energies) and duty cycles of these
instruments, and the details of the increasingly uncertain non-local
EBL at higher redshifts.  The field of extragalactic VHE astronomy has
grown considerably in the last five years, and ongoing progress on the
instrumentation front suggests that many new detections may be coming.
 
On Figure \ref{fig:attedge}, we have placed vertical lines at 50 and 100 GeV to facilitate comparison with the
gamma-ray attenuation edge curves.  In the fiducial WMAP5 model, the universe
does not become optically thick ($\tau > 1$) to 100 GeV gamma rays
until $z\sim 0.9$, and $z \sim 1.6$ for an observed energy of 50 GeV.
Our Fiducial model does predict slightly higher opacities at high
redshift compared to the fiducial model of \citet{gilmoreUV}; this is
due to more UV light escaping high-redshift galaxies in our evolving
dust model than in the fixed dust model used in the C\LCDM~prediction.
Nonetheless, the EBL does not become a significant barrier to VHE
observations at these low energies in either model until redshifts
considerably higher than those for which AGN have presently been
detected by ground-based instruments.  At the redshift of the current
most distant confirmed source, 3C279, we find $\tau=0.38$ for an
observed energy of 100 GeV, and $\tau=1$ at 175 GeV.

As mentioned, the approximation of a
local EBL in optical depth calculations is only valid for nearby
extragalactic observations.  At redshifts above $\sim$0.3, differences
in the evolution of star formation and galaxy emissivity begin to have
a substantial effect on attenuation; two different EBL models with the
same present-day normalization could have widely varying behavior at
these times.  For instance, the Kneiske models and Stecker's fast
evolution models have star-formation history peaks at a lower redshift
than our models predict.  In addition to predicting different results
for the present-day EBL than our model, the evolution with redshift is
also quite a bit different in these cases.  As no direct observations of the EBL are possible at nonzero redshift, predicting attenuation from sources past these distances must be made on the basis of models of galaxy evolution, constrained by surveys of luminosity functions at high redshift.  
Recent surveys of the non-local universe such as DEEP2 and the
multi-wavelength GOODS and AEGIS surveys have become powerful tools in
constraining the EBL at these distances.  Observational methods such
as \citet{franceschini08}, \citet{finke10}, and D11, which fall under
types (ii), (iii), and (iv) respectively in the classification scheme
discussed in the Introduction, are complementary to our semi-analytic
approach, which is of the first type.  Beyond $z\sim1$ to 2,
uncertainties in available star-formation and luminosity data become
large, and theoretical models will continue to be required to
understand the production and evolution of the background.

{The other impact of the shift to higher redshift observations by lower energy-threshold instruments such as the H.E.S.S.~phase-II upgrade and the future CTA experiment is the change in the relevant absorbing photon population to UV wavelengths.  Our models, and the others we reference in the previous section, predict a rapid falloff in transmission of gamma rays above 500 GeV for blazars at the redshift of 3C279.  Detecting emission at or above 1 TeV from sources at this distance will require orders-of-magnitude gains in instrument effective area, or observations of flare events with similar increases in output.  The energy range of primary interest for these types of sources is going to be 50 to 500 GeV for the next generation of IACTs, plus lower energy data from Fermi.  Below 200 GeV, it is the UV background that is primarily responsible for absorption.  In \citet{gilmoreUV}, we addressed the question of this background using four different models varying in high redshift star-formation and quasar emissivity; the `fiducial' model in that paper that we consider to be the most likely was based on the C\LCDM~model in this work.  The WMAP5 model with evolving dust parameters presented in this work is generally consistent with this model in its predictions for the evolution of UV emission, though it does predict a somewhat stronger UV background at high redshift.

One significant weakness of our present approach is our use of templates to describe re-emission by dust at mid- and far-IR wavelengths, which is relevant to the interpretation of multi-TeV data from nearby blazars.  This method makes the assumption that galaxies of a given bolometric luminosity emit light with a similar spectral distribution.  As discussed in SGPD11, future progress in understanding these wavelengths will likely require moving past this assumption.  While there is much progress to be made modeling this part of the background SED, new models of dust will only have a substantial effect on our calculation of gamma-ray opacities for the nearest VHE sources.

\section*{Acknowledgments}
AD, RCG, and JRP acknowledge support from a Fermi Guest Investigator
Grant.  RCG and JRP were also supported by NASA ATP grant NNX07AGG4G,
NSF-AST-1010033, and NSF-AST-0607712.  RCG was supported during this
work by a SISSA postdoctoral fellowship.  RSS and RCG thank UCSC for
hospitality during many visits.  A.D.'s work has been
supported by the Spanish Ministerio de Educaci\'on y Ciencia and the
European regional development fund (FEDER) under projects FIS2008-04189
and CPAN-Ingenio (CSD2007-00042), and by the Junta de Andaluc\'ia
(P07-FQM-02894).

We wish to thank Felix Aharonian, David A. Williams, Amy Furniss, and
A. Nepomuk Otte for helpful discussions concerning gamma--ray
attenuation.  Fabio Fontanot assisted in comparisons of different dust
templates.  We also thank the anonymous referee for comments which helped improve this paper

The TeVCat (http://tevcat.uchicago.edu/) online TeV-source catalogue
was used in the preparation of this paper.


\label{lastpage}
\end{document}